\pdfoutput=1

\documentclass[12pt,a4paper]{article}
\usepackage{overpic}
\usepackage{ifthen} 
\newboolean{pdflatex}
\setboolean{pdflatex}{true} 

\newboolean{articletitles}
\setboolean{articletitles}{true} 

\newboolean{uprightparticles}
\setboolean{uprightparticles}{false} 

\newboolean{inbibliography}
\setboolean{inbibliography}{false} 

\usepackage{microtype}
\usepackage{lineno}  
\usepackage{xspace} 
\usepackage{caption} 

\usepackage{graphicx}  
\usepackage{color}
\usepackage{colortbl}
\graphicspath{{./}} 

\usepackage{amsmath} 
\usepackage{amssymb}
\usepackage{amsfonts}
\usepackage{upgreek} 

\newcommand*\patchAmsMathEnvironmentForLineno[1]{%
\expandafter\let\csname old#1\expandafter\endcsname\csname #1\endcsname
\expandafter\let\csname oldend#1\expandafter\endcsname\csname
end#1\endcsname
 \renewenvironment{#1}%
   {\linenomath\csname old#1\endcsname}%
   {\csname oldend#1\endcsname\endlinenomath}%
}
\newcommand*\patchBothAmsMathEnvironmentsForLineno[1]{%
  \patchAmsMathEnvironmentForLineno{#1}%
  \patchAmsMathEnvironmentForLineno{#1*}%
}
\AtBeginDocument{%
\patchBothAmsMathEnvironmentsForLineno{equation}%
\patchBothAmsMathEnvironmentsForLineno{align}%
\patchBothAmsMathEnvironmentsForLineno{flalign}%
\patchBothAmsMathEnvironmentsForLineno{alignat}%
\patchBothAmsMathEnvironmentsForLineno{gather}%
\patchBothAmsMathEnvironmentsForLineno{multline}%
\patchBothAmsMathEnvironmentsForLineno{eqnarray}%
}

\usepackage{hyperref}    
\usepackage[all]{hypcap} 

\def\lhcb {\mbox{LHCb}\xspace}

\def\MagUp {\mbox{\em Mag\kern -0.05em Up}\xspace}

\ifthenelse{\boolean{uprightparticles}}%
{

 \def\Pmu         {\ensuremath{\upmu}\xspace}

 \def\Ppsi        {\ensuremath{\uppsi}\xspace}

 \def\PDelta      {\ensuremath{\Delta}\xspace}                 
 \def\PXi      {\ensuremath{\Xi}\xspace}                 
 \def\PLambda      {\ensuremath{\Lambda}\xspace}                 
 \def\PSigma      {\ensuremath{\Sigma}\xspace}                 
 \def\POmega      {\ensuremath{\Omega}\xspace}                 
 \def\PUpsilon      {\ensuremath{\Upsilon}\xspace}                 
 
 \def\PB      {\ensuremath{\mathrm{B}}\xspace}                 
                  
 \def\PD      {\ensuremath{\mathrm{D}}\xspace}

 \def\PJ      {\ensuremath{\mathrm{J}}\xspace}                 
 \def\PK      {\ensuremath{\mathrm{K}}\xspace}

 \def\Pb      {\ensuremath{\mathrm{b}}\xspace}                 
 \def\Pc      {\ensuremath{\mathrm{c}}\xspace}                 
 \def\Pd      {\ensuremath{\mathrm{d}}\xspace}

 \def\Pi      {\ensuremath{\mathrm{i}}\xspace}

 \def\Pp      {\ensuremath{\mathrm{p}}\xspace}

 \def\Ps      {\ensuremath{\mathrm{s}}\xspace}                 
                  
 \def\Pu      {\ensuremath{\mathrm{u}}\xspace}

}
{

 \def\Pmu         {\ensuremath{\mu}\xspace}

 \def\Ppsi        {\ensuremath{\psi}\xspace}                 
                  
 \mathchardef\PDelta="7101
 \mathchardef\PXi="7104
 \mathchardef\PLambda="7103
 \mathchardef\PSigma="7106
 \mathchardef\POmega="710A
 \mathchardef\PUpsilon="7107
                  
 \def\PB      {\ensuremath{B}\xspace}                 
                  
 \def\PD      {\ensuremath{D}\xspace}

 \def\PJ      {\ensuremath{J}\xspace}                 
 \def\PK      {\ensuremath{K}\xspace}

 \def\Pb      {\ensuremath{b}\xspace}                 
 \def\Pc      {\ensuremath{c}\xspace}                 
 \def\Pd      {\ensuremath{d}\xspace}

 \def\Pi      {\ensuremath{i}\xspace}

 \def\Pp      {\ensuremath{p}\xspace}

 \def\Ps      {\ensuremath{s}\xspace}                 
                  
 \def\Pu      {\ensuremath{u}\xspace}

}

\makeatletter
\ifcase \@ptsize \relax
  \newcommand{\miniscule}{\@setfontsize\miniscule{4}{5}}
\or
  \newcommand{\miniscule}{\@setfontsize\miniscule{5}{6}}
\or
  \newcommand{\miniscule}{\@setfontsize\miniscule{5}{6}}
\fi
\makeatother

\DeclareRobustCommand{\optbar}[1]{\shortstack{{\miniscule (\rule[.5ex]{1.25em}{.18mm})}
  \\ [-.7ex] $#1$}}

\def\mup        {{\ensuremath{\Pmu^+}}\xspace}
\def\mun        {{\ensuremath{\Pmu^-}}\xspace} 

\def\uquark    {{\ensuremath{\Pu}}\xspace}
\def\uquarkbar {{\ensuremath{\overline \uquark}}\xspace}

\def\dquark    {{\ensuremath{\Pd}}\xspace}

\def\squark    {{\ensuremath{\Ps}}\xspace}

\def\cquark    {{\ensuremath{\Pc}}\xspace}
\def\cquarkbar {{\ensuremath{\overline \cquark}}\xspace}

\def\bquark    {{\ensuremath{\Pb}}\xspace}

\def\kaon    {{\ensuremath{\PK}}\xspace}
  \def\Kbar    {{\kern 0.2em\overline{\kern -0.2em \PK}{}}\xspace}

\def\KorKbar    {\kern 0.18em\optbar{\kern -0.18em K}{}\xspace}

\def\Kp      {{\ensuremath{\kaon^+}}\xspace}

\def\Kstar   {{\ensuremath{\kaon^*}}\xspace}

  \def\Dbar    {{\kern 0.2em\overline{\kern -0.2em \PD}{}}\xspace}

\def\DorDbar    {\kern 0.18em\optbar{\kern -0.18em D}{}\xspace}

\def\B       {{\ensuremath{\PB}}\xspace}
\def\Bbar    {{\ensuremath{\kern 0.18em\overline{\kern -0.18em \PB}{}}}\xspace}

\def\BorBbar    {\kern 0.18em\optbar{\kern -0.18em B}{}\xspace}
\def\Bz      {{\ensuremath{\B^0}}\xspace}

\def\Bu      {{\ensuremath{\B^+}}\xspace}

\def\Bd      {{\ensuremath{\B^0}}\xspace}
\def\Bs      {{\ensuremath{\B^0_\squark}}\xspace}
\def\Bsb     {{\ensuremath{\Bbar{}^0_\squark}}\xspace}

\def\jpsi     {{\ensuremath{{\PJ\mskip -3mu/\mskip -2mu\Ppsi\mskip 2mu}}}\xspace}

  \def\Y#1S{\ensuremath{\PUpsilon{(#1S)}}\xspace}

\def\proton      {{\ensuremath{\Pp}}\xspace}

\def\Lz          {{\ensuremath{\PLambda}}\xspace}
\def\Lbar        {{\ensuremath{\kern 0.1em\overline{\kern -0.1em\PLambda}}}\xspace}
\def\LorLbar    {\kern 0.18em\optbar{\kern -0.18em \PLambda}{}\xspace}

\def\Lb      {{\ensuremath{\Lz^0_\bquark}}\xspace}

\def\to                 {\ensuremath{\rightarrow}\xspace}

\def\CP                {{\ensuremath{C\!P}}\xspace}

\def\AT#1     {\ensuremath{A_{\mathrm{T}}^{#1}}\xspace}           

\def\C#1      {\ensuremath{\mathcal{C}_{#1}}\xspace}                       
\def\Cp#1     {\ensuremath{\mathcal{C}_{#1}^{'}}\xspace}                    
\def\Ceff#1   {\ensuremath{\mathcal{C}_{#1}^{\mathrm{(eff)}}}\xspace}        
\def\Cpeff#1  {\ensuremath{\mathcal{C}_{#1}^{'\mathrm{(eff)}}}\xspace}       
\def\Ope#1    {\ensuremath{\mathcal{O}_{#1}}\xspace}                       
\def\Opep#1   {\ensuremath{\mathcal{O}_{#1}^{'}}\xspace}                    

\newcommand{\tev}{\ifthenelse{\boolean{inbibliography}}{\ensuremath{~T\kern -0.05em eV}\xspace}{\ensuremath{\mathrm{\,Te\kern -0.1em V}}}\xspace}
\newcommand{\gev}{\ensuremath{\mathrm{\,Ge\kern -0.1em V}}\xspace}
\newcommand{\mev}{\ensuremath{\mathrm{\,Me\kern -0.1em V}}\xspace}
\newcommand{\kev}{\ensuremath{\mathrm{\,ke\kern -0.1em V}}\xspace}
\newcommand{\ev}{\ensuremath{\mathrm{\,e\kern -0.1em V}}\xspace}
\newcommand{\gevc}{\ensuremath{{\mathrm{\,Ge\kern -0.1em V\!/}c}}\xspace}
\newcommand{\mevc}{\ensuremath{{\mathrm{\,Me\kern -0.1em V\!/}c}}\xspace}
\newcommand{\gevcc}{\ensuremath{{\mathrm{\,Ge\kern -0.1em V\!/}c^2}}\xspace}
\newcommand{\gevgevcccc}{\ensuremath{{\mathrm{\,Ge\kern -0.1em V^2\!/}c^4}}\xspace}
\newcommand{\mevcc}{\ensuremath{{\mathrm{\,Me\kern -0.1em V\!/}c^2}}\xspace}

\def\mum  {\ensuremath{{\,\upmu\rm m}}\xspace}

\def\invfb   {\ensuremath{\mbox{\,fb}^{-1}}\xspace}

\def\ps   {\ensuremath{{\rm \,ps}}\xspace}

\newcommand{\stat}{\ensuremath{\mathrm{\,(stat)}}\xspace}
\newcommand{\syst}{\ensuremath{\mathrm{\,(syst)}}\xspace}

\def\gsim{{~\raise.15em\hbox{$>$}\kern-.85em
          \lower.35em\hbox{$\sim$}~}\xspace}
\def\lsim{{~\raise.15em\hbox{$<$}\kern-.85em
          \lower.35em\hbox{$\sim$}~}\xspace}

\def\ptot       {\mbox{$p$}\xspace}
\def\pt         {\mbox{$p_{\rm T}$}\xspace}

\def\evtgen     {\mbox{\textsc{EvtGen}}\xspace}

\def\gauss      {\mbox{\textsc{Gauss}}\xspace}
\def\geant      {\mbox{\textsc{Geant4}}\xspace}

\def\photos     {\mbox{\textsc{Photos}}\xspace}

\def\pythia     {\mbox{\textsc{Pythia}}\xspace}

\def\tell1  {TELL1\xspace}
\def\ukl1   {UKL1\xspace}

\def\cospsi{\cos\theta_{\mu}}
\def\cosks{\cos\theta_{K}}

\def\Bsbd     {{\ensuremath{\B^0_{(\squark)}}}\xspace}

\usepackage{cite} 
\usepackage{mciteplus}
\usepackage{longtable} 

\begin{document}

\renewcommand{\thefootnote}{\fnsymbol{footnote}}
\setcounter{footnote}{1}

\begin{titlepage}
\pagenumbering{roman}

\vspace*{-1.5cm}
\centerline{\large EUROPEAN ORGANIZATION FOR NUCLEAR RESEARCH (CERN)}
\vspace*{1.5cm}
\hspace*{-0.5cm}
\begin{tabular*}{\linewidth}{lc@{\extracolsep{\fill}}r}
\ifthenelse{\boolean{pdflatex}}
{\vspace*{-2.7cm}\mbox{\!\!\!\includegraphics[width=.14\textwidth]{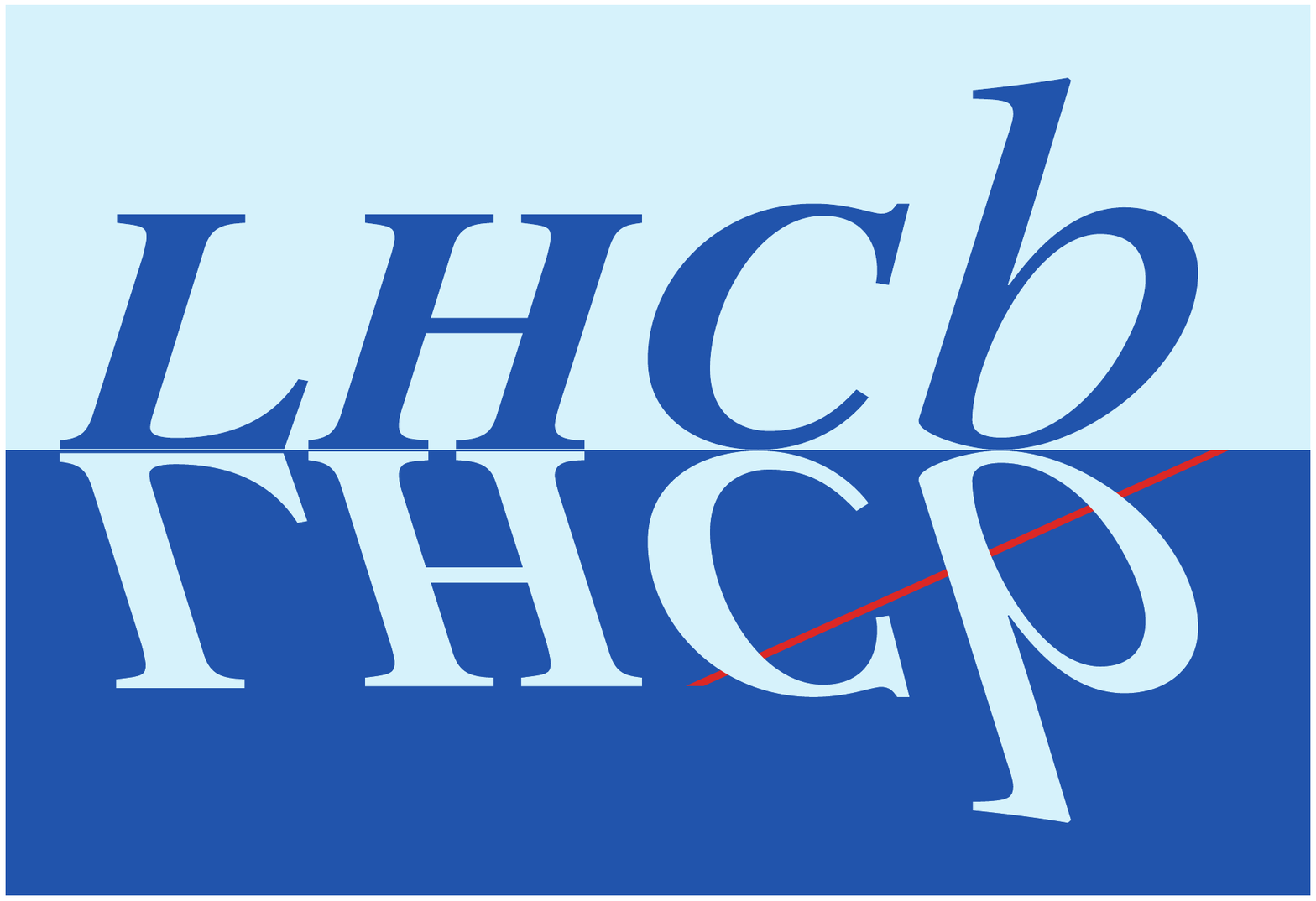}} & &}%
{\vspace*{-1.2cm}\mbox{\!\!\!\includegraphics[width=.12\textwidth]{lhcb-logo.eps}} & &}%
\\
 & & CERN-PH-EP-2015-072 \\  
 & & LHCb-PAPER-2015-010 \\  
 & & June 12, 2015 \\ 
 & & \\
\end{tabular*}

\vspace*{2.0cm}

{\bf\boldmath\huge
\begin{center}
Observation of the decay $\overline{B}_s^0 \rightarrow \psi(2S) K^+ \pi^-$
\end{center}
}

\vspace*{0.5cm}

\begin{center}
The LHCb collaboration\footnote{Authors are listed at the end of this paper.}
\end{center}

\vspace{\fill}

\begin{abstract}
  \noindent
The decay $\overline{B}_s^0 \rightarrow \psi(2S) K^+ \pi^-$ is
observed using a data set corresponding to an integrated luminosity of $3.0 \invfb$
collected by the LHCb experiment in $pp$ collisions at
centre-of-mass energies of 7 and 8 TeV. The branching fraction relative to the $\Bz\rightarrow \psi(2S)
K^+ \pi^-$ decay mode is measured to be
\begin{equation}
\frac{{\cal B}(\overline{B}^0_s \rightarrow \psi(2S) K^+ \pi^-)}{{\cal B}(B^0
  \rightarrow \psi(2S) K^+ \pi^-)} = 5.38 \pm  0.36 \stat \pm 0.22 \syst
\pm 0.31  \, (f_s/f_d ) \, \%,\nonumber
\end{equation}
where $f_s/f_d$ indicates the uncertainty due to the ratio of
probabilities for a $b$ quark to hadronise into a $\Bs$ or $\Bz$ meson. Using an amplitude
analysis, the fraction of decays proceeding via an intermediate $K^*(892)^0$ meson
is measured to be $0.645 \pm 0.049 \stat \pm 0.049 \syst$ and
its longitudinal polarisation fraction
is $0.524 \pm 0.056 \stat \pm 0.029 \syst$. 
The relative branching
fraction for this component is determined to be
\begin{equation}
\frac{{\cal B}(\overline{B}^0_s \rightarrow \psi(2S) K^*(892)^0)}{{\cal B}(B^0
  \rightarrow \psi(2S) K^*(892)^0)} = 5.58 \pm  0.57 \stat \pm 0.40 \syst
\pm 0.32  \, (f_s/f_d)  \, \%. \nonumber
\end{equation}
In addition, the mass splitting between the $\Bs$ and $\Bz$ mesons is measured as
\begin{equation}
M(B^0_s) - M(B^0)  = 87.45 \pm 0.44  \stat \pm 0.09 \syst
\mevcc. \nonumber
\end{equation}
  
\end{abstract}

\vspace*{0.50cm}

\begin{center}
Published in Phys.~Lett.~B 
\end{center}

\vspace{\fill}

{\footnotesize 
\centerline{\copyright~CERN on behalf of the \lhcb collaboration, license \href{http://creativecommons.org/licenses/by/4.0/}{CC-BY-4.0}.}}
\vspace*{1mm}

\end{titlepage}


\newpage
\setcounter{page}{2}
\mbox{~}

\cleardoublepage

\renewcommand{\thefootnote}{\arabic{footnote}}
\setcounter{footnote}{0}



\pagestyle{plain} 
\setcounter{page}{1}
\pagenumbering{arabic}


%

\newpage

\section{Introduction}
\label{sec:Introduction}
\begin{figure}[hb!]
\centering
\includegraphics[width=0.45\textwidth]{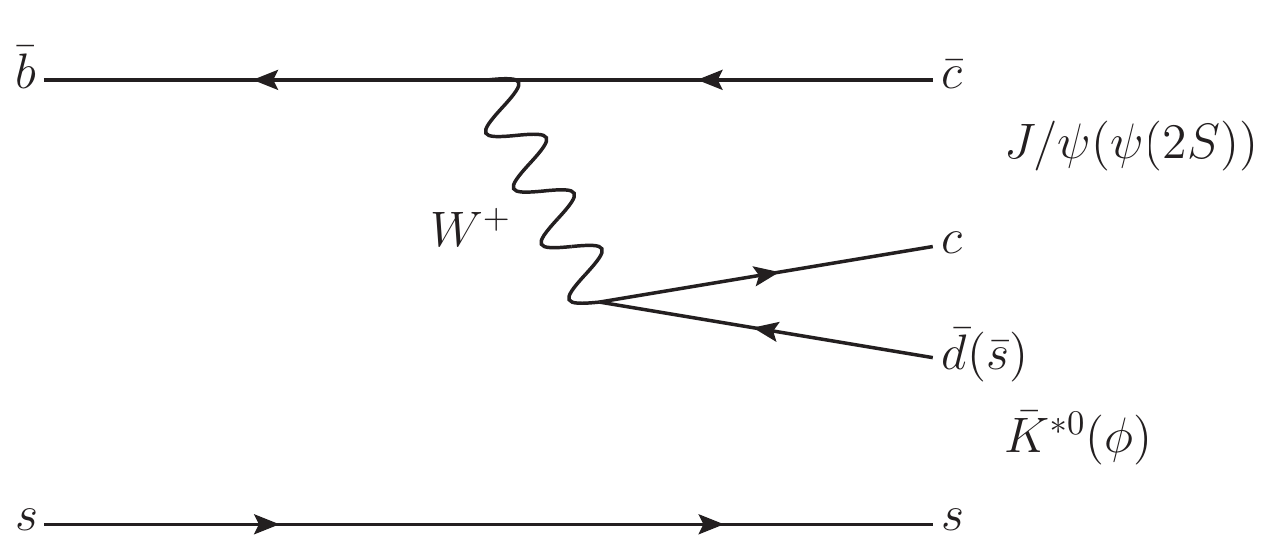}
\includegraphics[width=0.45\textwidth]{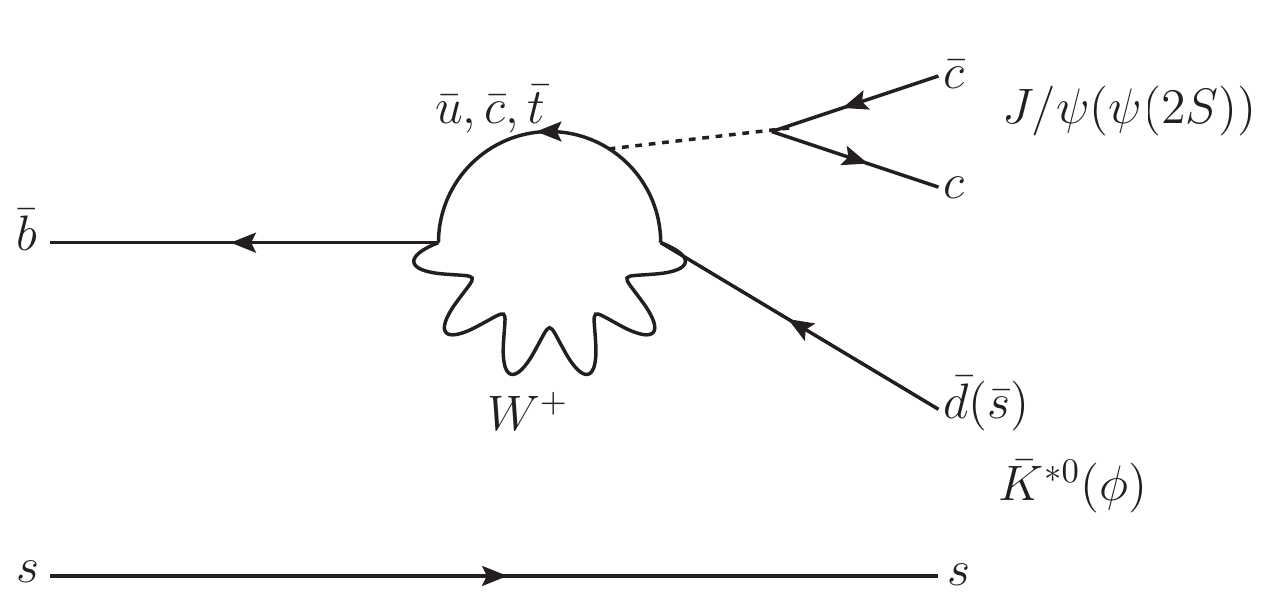}
\caption{\small
Tree (left) and penguin (right) topologies contributing to  
the $\Bsbd \rightarrow \psi V$ decays where $\psi = J/\psi, 
\psi(2S)$ and $V = \phi, K^*(892)^0$.
}
\label{fig:feynman}
\end{figure}

The large data set collected by the LHCb experiment has allowed
precision measurements of time-dependent \CP violation 
in the $\Bs \rightarrow J/\psi \phi$ and $\Bs \rightarrow
J/\psi f_0(980)$ decay modes \cite{LHCb-PAPER-2014-059,LHCb-PAPER-2012-006}.\footnote{Charge-conjugatation is implicit unless stated otherwise.}
The results are interpreted assuming that these
decays are dominated by colour-suppressed tree-level amplitudes 
(Fig.~\ref{fig:feynman}). Higher-order 
penguin amplitudes, which are difficult to calculate in QCD, also
contribute (Fig.~\ref{fig:feynman}). Reference \cite{Faller:2008gt} suggests that the
size of contributions from these processes can be determined by studying
decay modes such as $\Bsb \rightarrow J/\psi K^*(892)^0$
where they dominate.
The $\Bsb \rightarrow J/\psi K^*(892)^0$ decay mode was first observed by the CDF collaboration \cite{Aaltonen:2011sy}
and subsequently studied in detail by the LHCb collaboration \cite{LHCb-PAPER-2012-014}.

Recently, interest in $b$-hadron decays to final states containing
charmonia has been generated by the observation of
the $Z(4430)^-\to \psi(2S)\pi^-$ state in the $\Bz \rightarrow \psi(2S) K^+ \pi^-$ decay
chain by the
Belle \cite{Choi:2007wga,Mizuk:2009da,Chilikin:2013tch} and LHCb
collaborations \cite{LHCb-PAPER-2014-014}. As this state is charged
and has a minimal
quark content of $\cquark \cquarkbar \dquark \uquarkbar$, it 
is interpreted as evidence for the existence of non-$q\overline{q}$
mesons \cite{Klempt:2007cp}. Evidence for similar exotic structures in $\Bz
\rightarrow \chi_{c1, c2} K^+ \pi^-$ and   $\Bz
\rightarrow J/\psi K^+ \pi^-$decays has been reported by the Belle
collaboration\cite{Mizuk:2008me, Chilikin:2014bkk}. If these structures correspond to
real particles they should be visible in other decay modes.

This letter reports the first observation of the decay
$\overline{B}_s^0 \rightarrow \psi(2S)K^+\pi^-$ and presents measurements
of the inclusive branching fraction and the fraction of decays that proceed via 
an intermediate $\Kstar(892)^0$ resonance, as determined from an amplitude
analysis of the final state. The amplitude analysis also allows the determination
of the longitudinal polarisation fraction of the $\Kstar(892)^0$ meson.
Additionally a measurement of the mass difference between \Bs and \Bd
mesons is reported that improves the current knowledge of this observable.

\section{Detector and simulation}
\label{sec:Detector}
 The \lhcb detector~\cite{Alves:2008zz,LHCb-DP-2014-002} is a single-arm forward
 spectrometer covering the \mbox{pseudorapidity} range $2<\eta <5$,
 designed for the study of particles containing \bquark or \cquark
 quarks. The detector includes a high-precision tracking system
 consisting of a silicon-strip vertex detector surrounding the $pp$
 interaction region, a large-area silicon-strip detector located
 upstream of a dipole magnet with a bending power of about
 $4{\rm\,Tm}$, and three stations of silicon-strip detectors and straw
 drift tubes~\cite{LHCb-DP-2013-003} placed downstream of the magnet.
 The tracking system provides a measurement of momentum, \ptot, of charged particles with
 a relative uncertainty that varies from 0.5\% at low momentum to 1.0\% at 200\gevc.
 The minimum distance of a track to a primary vertex, the impact parameter, is measured with a resolution of $(15+29/\pt)\mum$,
 where \pt is the component of the momentum transverse to the beam, in\,\gevc.
 Large samples of 
$B^+ \to \jpsi K^+$ and $\jpsi\to\mup\mun$ decays, collected 
concurrently with the data set used here, were used to calibrate the
momentum scale of the spectrometer to a precision of $0.03\,\%$
\cite{LHCb-PAPER-2012-048}. 

 Different types of charged hadrons are distinguished using information
 from two ring-imaging Cherenkov detectors~\cite{LHCb-DP-2012-003}.
 Photons, electrons and hadrons are identified by a calorimeter system consisting of
 scintillating-pad and preshower detectors, an electromagnetic
 calorimeter and a hadronic calorimeter. Muons are identified by a
 system composed of alternating layers of iron and multiwire
 proportional chambers~\cite{LHCb-DP-2012-002}.
 The online event selection is performed by a trigger~\cite{LHCb-DP-2012-004},
 which consists of a hardware stage, based on information from the calorimeter and muon
 systems, followed by a software stage, which applies a full event
 reconstruction. In this analysis candidates are first required to pass the
hardware trigger, which selects muons and dimuon pairs based on the
transverse momentum. At the subsequent software stage, events 
are triggered by a $\psi(2S) \rightarrow \mu^+ \mu^-$ candidate
where the $\psi(2S)$ is required to be consistent with coming from 
the decay of a $\bquark$ hadron by either using impact parameter requirements
on daughter tracks
or detachment of the $\psi(2S)$ candidate from the primary vertex.

The analysis is performed using data corresponding to an integrated
luminosity of 1.0\,fb$^{-1}$ collected in $pp$ collisions at a 
centre-of-mass energy of 7~TeV and 2.0\,fb$^{-1}$
collected at 8~TeV. In the simulation, $pp$ collisions are generated using
\pythia~\cite{Sjostrand:2006za,*Sjostrand:2007gs}  with a specific \lhcb
configuration~\cite{LHCb-PROC-2010-056}.  Decays of hadronic particles
are described by \evtgen~\cite{Lange:2001uf}, in which final state
radiation is generated using \photos~\cite{Golonka:2005pn}. The
interaction of the generated particles with the detector and its
response are implemented using the \geant 
toolkit~\cite{Allison:2006ve, *Agostinelli:2002hh} as described in
Ref.~\cite{LHCb-PROC-2011-006}.
\section{Event selection}
\label{sec:selection}
The selection of candidates is divided into two
parts. First, a loose selection is performed that
retains the majority of signal events whilst reducing the background
substantially. After this the $\Bd \rightarrow \psi(2S) K^+ \pi^-$ peak
is clearly visible. Subsequently, a multivariate method is used to further improve
the signal-to-background ratio and to
allow the observation of the $\Bsb \rightarrow \psi(2S) K^+ \pi^-$
decay.

The selection starts by reconstructing the dimuon decay of the
$\psi(2S)$ meson. Pairs of oppositely charged particles
identified as muons with $\pt > 550 \mevc$ are combined to form 
$\psi(2S)$ candidates. The invariant mass of the dimuon pair is
required to be within $60 \mevcc$ of the known $\psi(2S)$ mass
\cite{PDG2014}. To form $\Bsbd$ candidates, the selected $\psi(2S)$ 
mesons are combined with oppositely charged kaon and pion candidates.
Tracks that do not correspond to actual trajectories of charged particles are
suppressed by requiring that they have $\pt > 250 \mevc$ and
by selecting on the output of a neural
network trained to discriminate between these and genuine tracks associated to particles.
Combinatorial
background from hadrons originating in the primary vertex (PV) is suppressed
by requiring that both hadrons are significantly displaced from any PV.
Well-identified hadrons are selected using the information 
provided by the Cherenkov detectors. This is combined with kinematic
information using a neural network to provide a probability that a
particle is a kaon ($\mathcal{P}^K$), pion
($\mathcal{P}^{\pi}$) or proton ($\mathcal{P}^{p}$). It is required that $\mathcal{P}^K$ is larger
than 0.1 for the $K^+$ candidate and that $\mathcal{P}^{\pi}$ is larger than 0.2 for the $\pi^-$ candidate. 

A kinematical vertex fit is applied to the $\Bsbd$
candidates~\cite{Hulsbergen:2005pu}. 
To improve the invariant mass resolution, the fit is performed with
the requirement that the $\Bsbd$ candidate points to the PV and the
$\psi(2S)$ is mass constrained to the known value \cite{PDG2014}. 
A good quality of the vertex fit $\chi^2$, $\chi^2_{\mathrm{DTF}}$, is required.
To ensure good separation between the
$\Bz$ and $\Bs$ signals, the uncertainty on the reconstructed mass
returned by the fit must be less than 11\mevcc.
Combinatorial background from particles produced in the primary
vertex is further reduced by requiring the decay time of the
$\Bsbd$~meson to exceed $0.3\ps$. 

Four criteria are applied to reduce background from specific $\bquark$-hadron
decay modes. 
First, the candidate is rejected if the invariant mass of the hadron pair  
calculated assuming that both particles are kaons is within $10 
\mevcc$ of the known $\phi$ meson mass \cite{PDG2014}, 
suppressing $\Bs \rightarrow \psi(2S) \phi$ decays where one of the kaons is
misidentified as a pion. Second, to suppress $B^0
\rightarrow \psi(2S) \pi^+ \pi^-$ events where one of the pions is incorrectly identified as a kaon, it is required that
$\mathcal{P}^K > \mathcal{P}^{\pi}$ for the kaon candidate. This
rejects $80 \, \%$ of the background from this source whilst retaining
$90 \, \%$ of $\Bsbd$ signal candidates. Third, to suppress
background from $\Lb \rightarrow \psi(2S) \proton \pi^-$ decays where the proton
is misidentifed as a kaon,
candidates with $\mathcal{P}^p > 0.3$ and
an invariant mass within $15 \mevcc$ of the known $\Lb$ mass
\cite{PDG2014} are
discarded. Finally, to reduce background from a $B^+
\rightarrow \psi(2S) K^+$ decay combined with a random pion,
candidates where the reconstructed $\psi(2S) K^+$ invariant mass is
within $16 \mevcc$ of the known $B^+$ mass \cite{PDG2014} are
rejected. Background from the decay $\Lb \rightarrow \psi(2S)
\proton K^-$ with misidentified hadrons does not peak at the $\Bs$ mass and
is modelled in the fit.

To further improve the signal-to-background ratio, a multivariate
analysis based on a neural network is used. This is trained using simulated $\Bz$ signal events together
with candidates from data with a mass between $5500$ and $5600
\mevcc$ that are not used for subsequent analysis. Eight variables that give good separation between signal
and background are used: the number of clusters
in the large-area silicon tracker upstream of the magnet, 
$\mathcal{P}^{K}$ for the kaon candidate, $\mathcal{P}^{\pi}$ for 
the pion candidate, the transverse momentum of the $\Bsbd$, 
the minimum impact parameter to any primary vertex for each of the two
hadrons, $\chi^2_{\mathrm{DTF}}$ and the flight distance in the laboratory frame of
the $\Bsbd$ candidate divided by its uncertainty. The ratio $N_{\rm S}/\sqrt{N_{\rm S}+N_{\rm B}}$ is
used as a figure of merit, where $N_{\rm S}\ (N_{\rm B})$ is the number of signal (background) events
determined from the invariant mass fit (see Sect.~\ref{sec:massfit}).
The maximum value of this ratio is found for a threshold on the
neural network output that rejects $98\%$ of the background and retains $81\%$
of the signal for subsequent analysis. 

\section{Invariant mass fit}
\label{sec:massfit}
A maximum likelihood fit is made to the unbinned $\psi(2S) K^+ \pi^-$
invariant mass distribution, $m(\psi(2S)\Kp\pi^-)$, to extract the $\Bz$ and
$\Bs$ signal yields. The $\Bz$ signal component is
modelled by the sum of two Crystal Ball functions
\cite{Skwarnicki:1986xj} with common tail parameters and an additional Gaussian
component, all with a common mean. All parameters
are fixed to values determined from the simulation
apart from the common mean and an overall resolution scale
factor. The simulation is tuned to match the invariant mass
resolution seen in data for the $\Bu \rightarrow J/\psi K^+$ and $\Bd\to\jpsi \Kp\pi^-$ decay
modes. Consequently, the resolution scale factor is consistent
with unity in the fit to data. The $\Bs$ component is modelled with the same
function, with the mean value of the $\Bs$ meson mass left free in the fit.
The resolution parameters in this case are multiplied by a
factor of 1.06, determined from simulation, which accounts for the
additional energy release in this decay. 

The dominant background is combinatorial and modelled by an exponential
function. A significant component from $\Bs \rightarrow \psi(2S) \phi$
decays is visible at lower masses than the $\Bz$ peak. This is modelled
in the fit by a bifurcated Gaussian function where the shape
parameters are constrained to the values obtained in the simulation and
the yield constrained to the value determined in data under the hypothesis that both hadrons
are kaons. Additional small components
from $B^0_{(s)} \rightarrow \psi(2S) \pi^+ \pi^-$ and $\Lb \rightarrow \psi(2S)
\proton K^-$ decays are modelled by bifurcated Gaussian functions. The shapes of
these components are fixed using the simulation and the yields are
determined by normalising the simulation samples to the number of
candidates for each modes found in data using dedicated
selections.  Contributions from partially reconstructed decays are
accounted for in the combinatorial background.
In total, the fit has ten free
parameters. Variations of this fit model are considered as systematic uncertainties.

\begin{figure}[htb!]
\begin{center}
\resizebox{4.8in}{!}{\includegraphics{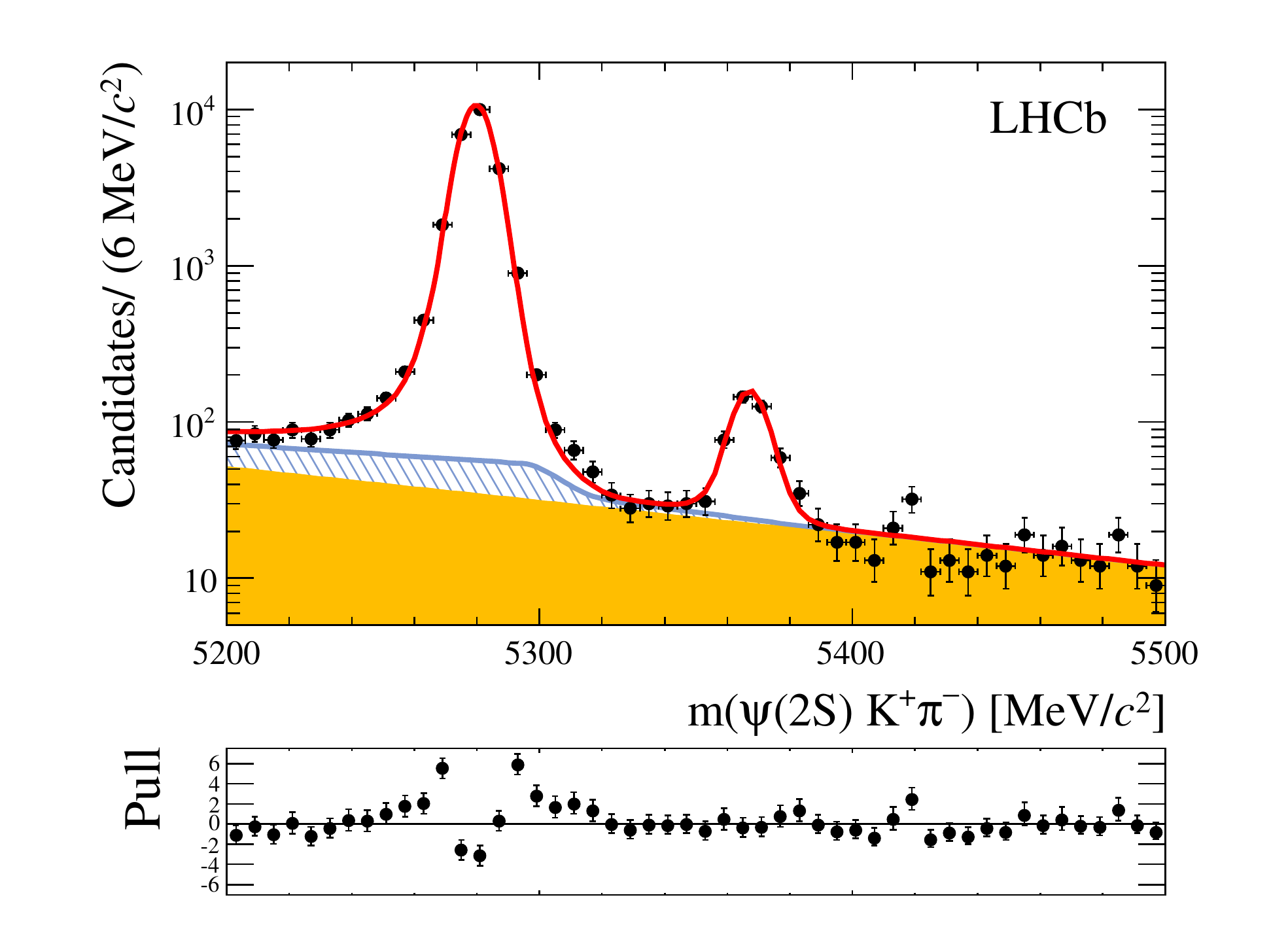}}
\caption{\small Invariant mass distribution for selected $\psi(2S)K^+ \pi^-$ candidates in
the data. A fit to the model described in the text is superimposed. The
full fit model is shown by the solid (red) line, the combinatorial
background by the solid (yellow) and the sum of background from the
exclusive $b \rightarrow \psi(2S)X$ modes considered in the text by the shaded (blue) area.
The maximum of
the $y$-scale is restricted so as to be able to see more clearly the 
$\Bsb \rightarrow \psi(2S)K^+ \pi^-$ signal. The lower plot shows the differences between the fit and measured
values divided by the corresponding uncertainty of the measured value,
the so-called pull distribution.}
\label{fig:massplot}
\end{center}
\end{figure}

Figure~\ref{fig:massplot} shows the invariant mass distribution observed
in the data together with the result of a fit to the model described
above. Binning the data,  a $\chi^2$-probability of
0.30 is found. The moderate mismodelling of the \Bd peak is
accounted for in the systematic uncertainties. The fit determines that there are
$329 \pm 22$ $\Bs$ decays and 
$24207 \pm 160$ $\Bz$ decays. The $\Bsb \rightarrow \psi(2S) K^+
\pi^-$ mode is observed with high significance.

The precision of the momentum scale calibration of $0.03\%$
translates to an uncertainty on the \Bd and \Bs meson masses of $0.3 
\mevcc$. Therefore, it is chosen to quote only the mass difference in
which this uncertainty largely cancels,
\begin{equation}
M(\Bs) -M(\Bz)  = 87.45 \pm 0.44\stat \pm 0.09\syst \mevcc. \nonumber
\end{equation} 
This procedure has been checked using the simulation,
which gives the input mass difference with a bias of $0.05 \mevcc$
that is assigned
as a systematic uncertainty. Further systematic uncertainties arise
from the momentum scale and mass fit model. Varying
the momentum scale by $0.03\%$ leads to an uncertainty of $0.04 \mevcc$.
The effect of the fit model is evaluated by considering several variations:
the relative fraction of the two Crystal Ball
functions is left free; the slope of the combinatorial background is
constrained using candidates where the kaon and pion have the
same charge; the Gaussian constraints on the background from the $\Bs
\rightarrow \psi(2S) \phi$ mode are removed; and the tail parameters of
the Crystal Ball functions are left free. The largest variation in the
mass splitting is $0.06 \, \mevcc$. 
The total systematic uncertainty is given by summing the individual
components in quadrature.

\section{Amplitude analysis}
\label{sec:angular}

\begin{figure}[t]
  \begin{center}
    \includegraphics*[scale=0.5]{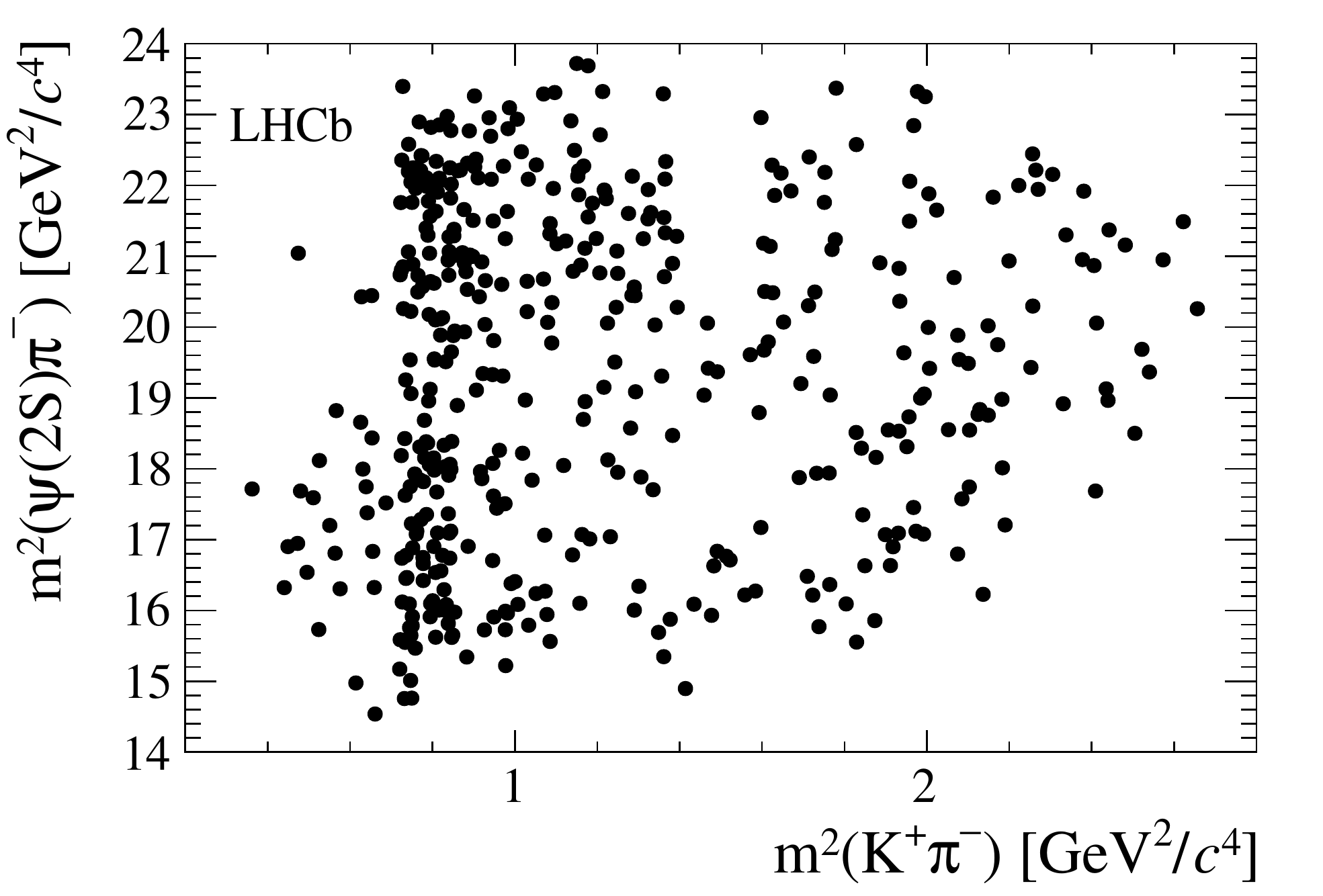} 
  \end{center}
  \vspace{-0.3cm}\caption{\small 
     Dalitz plot for the selected $\Bsb\to\psi(2S)\Kp\pi^-$  candidates
     in the signal window $m(\psi(2S)\Kp\pi^-) \in [5350, 5380]\mevcc$.
     \label{fig:dalitz}
  }
\end{figure}

\begin{figure}[t]
  \begin{center}
    \includegraphics[scale=0.2]{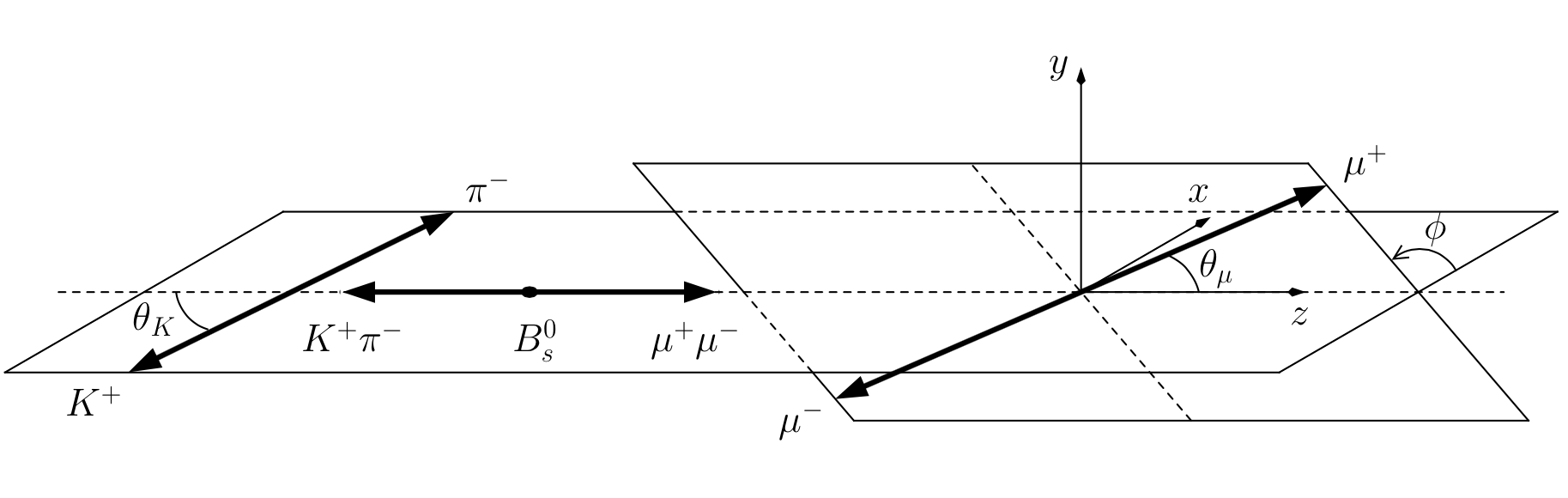} 
  \end{center}
  \vspace{-0.3cm}\caption{\small 
     Definition of the helicity angles.
 }    
   \label{fig:hel_angles}
\end{figure}

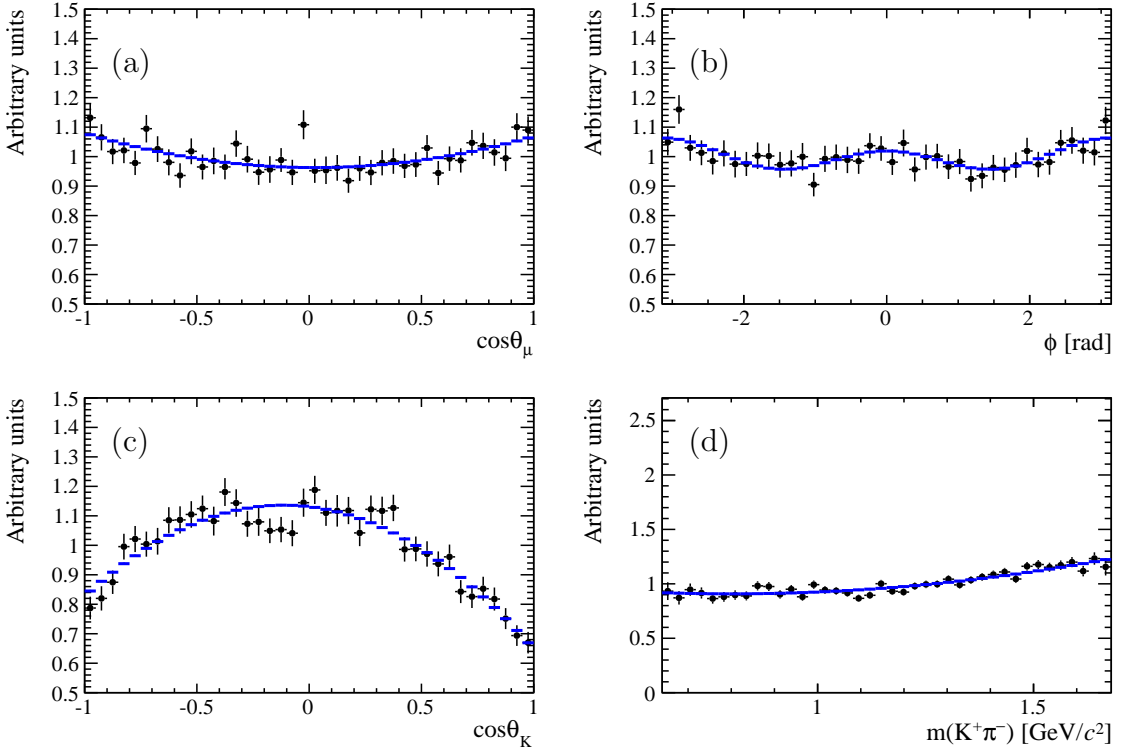
\begin{figure}[t]
  \begin{center}
  \begin{overpic}[width=0.95\textwidth]{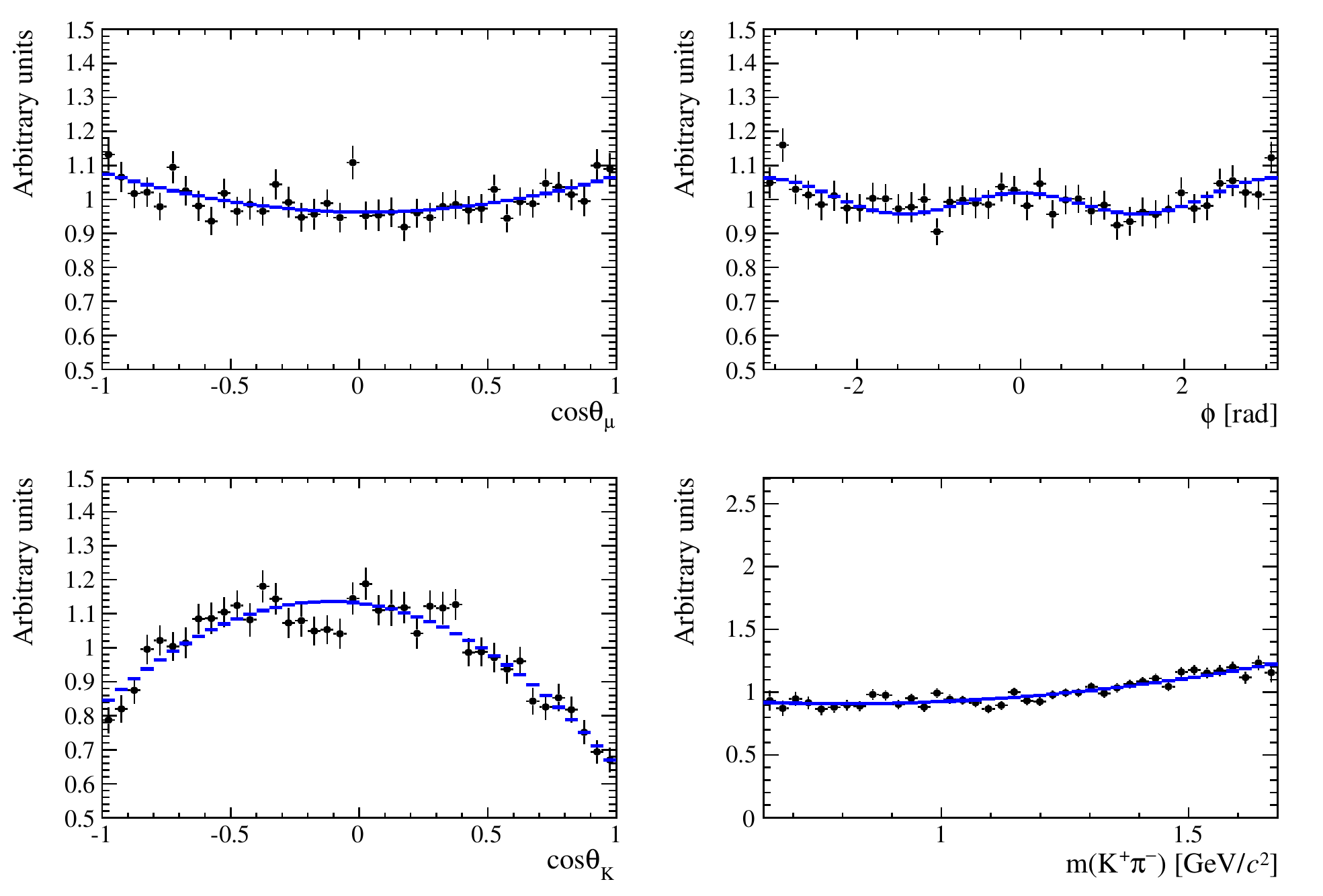} 
  \put(10,60){(a)}
  \put(60,60){(b)}
  \put(10,27){(c)}
  \put(60,27){(d)}
  \end{overpic}
  \end{center}
  \vspace{-0.3cm}\caption{\small 
  Distributions of (a) $\cos\theta_\mu$, (b) $\phi$, (c) $\cos\theta_K$ and (d) $m(\Kp\pi^-)$ 
     of simulated $B^0_s\to\psi(2S)\Kp\pi^-$ decays in a phase space configuration (black points)
     with the parameterisation of the efficiency overlaid (blue lines).
     }
  \label{fig:acceptance}
 
\end{figure}

Figure~\ref{fig:dalitz}
shows the Dalitz plot of the selected $\Bsb\to\psi(2S)\Kp\pi^-$ candidates
in the signal range, $m(\psi(2S)\Kp\pi^-) \in [5350, 5380]\mevcc$.
There is a clear enhancement around the known $\Kstar(892)^0$ mass~\cite{PDG2014}
and no other significant enhancements elsewhere.
To determine the fraction of decays that proceed via the $\Kstar(892)^0$
resonance, an amplitude analysis is performed, similar to that
used in Ref.~\cite{LHCb-PAPER-2014-014} for the analysis of the
$\Bz \rightarrow \psi(2S) K^+ \pi^-$ mode.
The final-state particles are
described using three angles $\Omega \equiv (\cos\theta_K, \cos\theta_\mu, \phi)$
in the helicity basis, defined in Fig.~\ref{fig:hel_angles},
and the invariant $K^+ \pi^-$ mass, $m_{K\pi} \equiv m(\Kp\pi^-)$.
The total amplitude is $S(m_{K\pi}, \Omega)\varepsilon(m_{K\pi}, \Omega) + B(m_{K\pi}, \Omega)$, 
where $S(m_{K\pi}, \Omega)$ represents the coherent sum over the helicity amplitudes for each
considered $\Kp\pi^-$ resonance or non-resonant component. 
The detection efficiency, $\varepsilon(m_{K\pi},\cosks,\cospsi,\phi)$, is evaluated
using simulation and parameterised using a combination of Legendre
polynomials and spherical harmonic moments, given by
\begin{equation}\small
\varepsilon(m_{K\pi},\cosks,\cospsi,\phi) =
\sum_{a,b,c,d} c^{abcd} P_a(\cosks) Y_{bc}(\cospsi,\phi) P_d\left( \frac{2(m_{K\pi}-m_{K\pi}^{\rm min})}{m_{K\pi}^{\rm max}-m_{K\pi}^{\rm min}}-1 \right),
\end{equation}
where $m_{K\pi}^{\rm min(max)}$ is the minimum (maximum) value allowed for
$m_{K\pi}$ in the available phase space of the decay.
The coefficients of the efficiency parameterisation
are computed by summing over the $N_{\rm MC}$ events simulated uniformly in the phase-space as
\begin{equation}
c^{abcd} = \frac{1}{N_{\rm MC}} \sum_i^{N_{\rm MC}} \frac{2a+1}{2} \frac{2d+1}{2} 
P_a({\cosks}_i) Y_{bc}({\cospsi}_i,{\phi}_i) P_d\left( \frac{2({m_{K\pi}}_i-m_{K\pi}^{\rm min})}{m_{K\pi}^{\rm max}-m_{K\pi}^{\rm min}}-1 \right) 
\frac{C}{g_i},
\label{eq:effmoments}
\end{equation}
where $g_i=p_iq_i$, with $p_i\, (q_i)$ being the  momentum of the $K^+\pi^-$ system
($K^+$ meson) in the $B^0\ (K^+\pi^-)$ rest frame and $C$ is a normalising constant with units ${\rm GeV}^2/c^2$. 
This approach provides a description of the multidimensional correlations without assuming factorisation.
In practice, the sum is over a finite number of moments ($a\le2$, $b\le2$, $c\le2$ and $d\le2$)
and only coefficients with a statistical significance larger than five standard deviations
from zero are retained. The one-dimensional projections of the parameterised efficiency are shown in 
Fig.~\ref{fig:acceptance}, superimposed on the simulated event distributions.

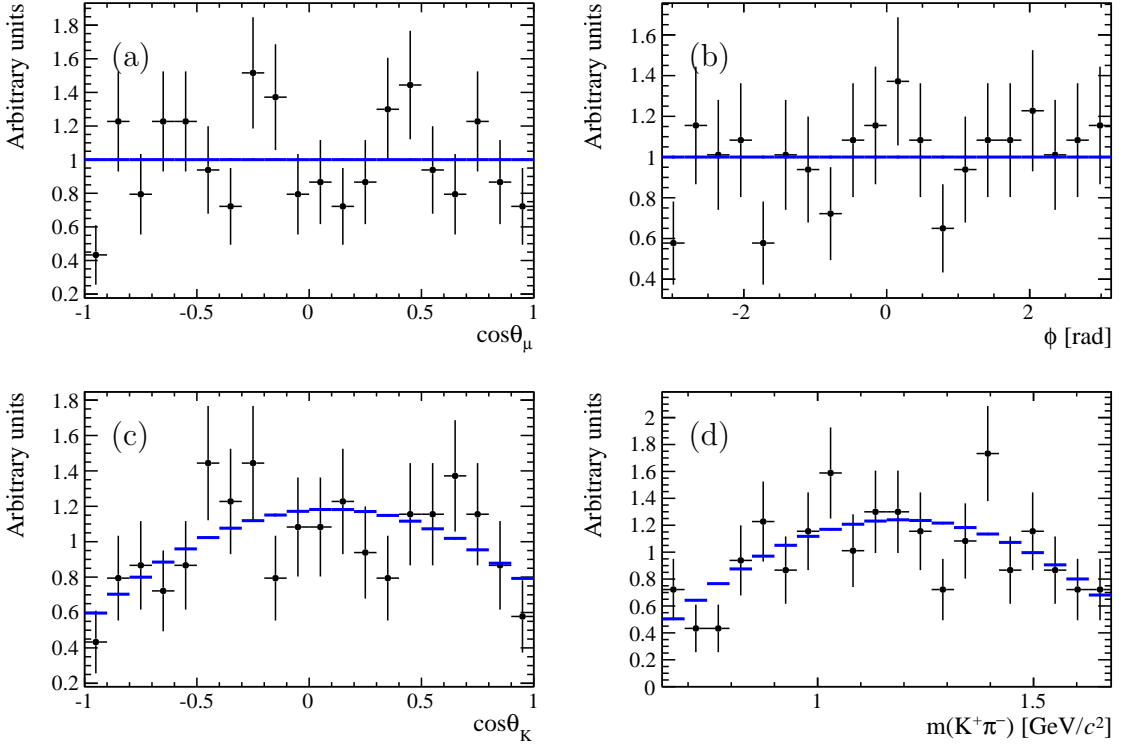
\begin{figure}[t]
  \begin{center}
    \begin{overpic}[width=0.95\textwidth]{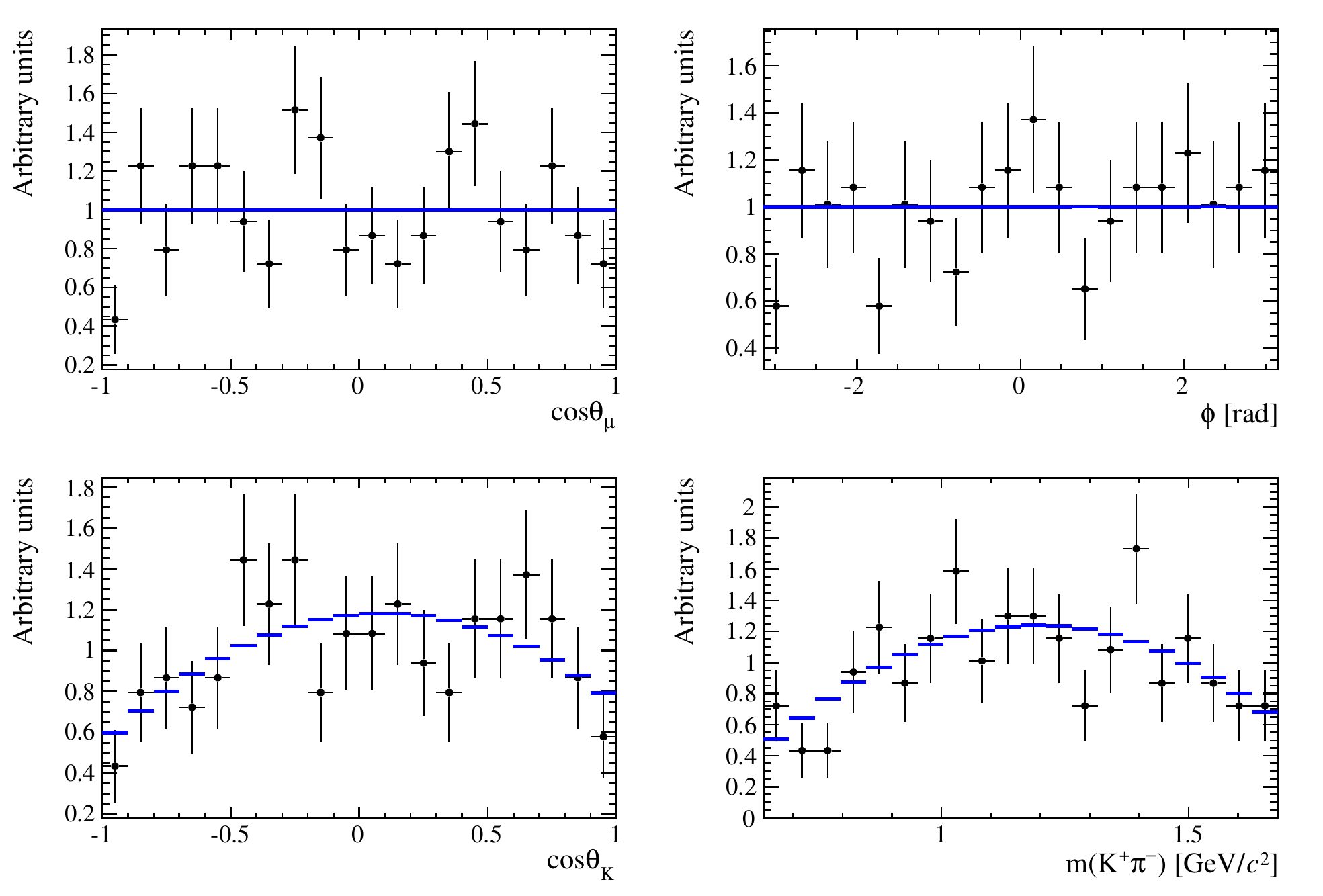} 
  \put(10,60){(a)}
  \put(60,60){(b)}
  \put(10,27){(c)}
  \put(60,27){(d)}
  \end{overpic}
  \end{center}
  \vspace{-0.3cm}\caption{\small 
     Distributions of (a) $\cos\theta_\mu$, (b) $\phi$, (c) $\cos\theta_K$ and (d) $m(\Kp\pi^-)$ 
     of $B^0_{(s)}\to\psi(2S)\Kp\pi^-$ candidates with $m(\psi(2S)\Kp\pi^-) > 5390\mevcc$ (black points),
     with the parameterisation of the background distribution overlaid (blue lines).
  }
 \label{fig:backgrounds}
\end{figure}

The background probability density function, $B(m_{K\pi}, \Omega)$, is determined using
a similar method as for the efficiency parameterisation. In this case the
sum in Eq.~(\ref{eq:effmoments}) is over the selected events with
$m(\psi(2S)\Kp\pi^-)>5390\mevcc$ and $g_i\equiv 1$.
Only moments with $a\le2$, $b=0$, $c=0$ and $d\le2$ and a statistical significance
larger than five standard deviations from zero are retained.
The one-dimensional projections of the parameterised background distribution are
shown in Fig.~\ref{fig:backgrounds},  superimposed on the sideband data. 
As a consistency check, the $(m_{K\pi}, \Omega)$ distributions for events with
$m(\psi(2S)\Kp\pi^-)>5390\mevcc$ are found to be compatible with the same distributions
obtained from a like-sign ($\psi(2S)K^{\pm}\pi^{\pm}$) sample.

\begin{figure}[t]
    \begin{overpic}[scale=0.41]{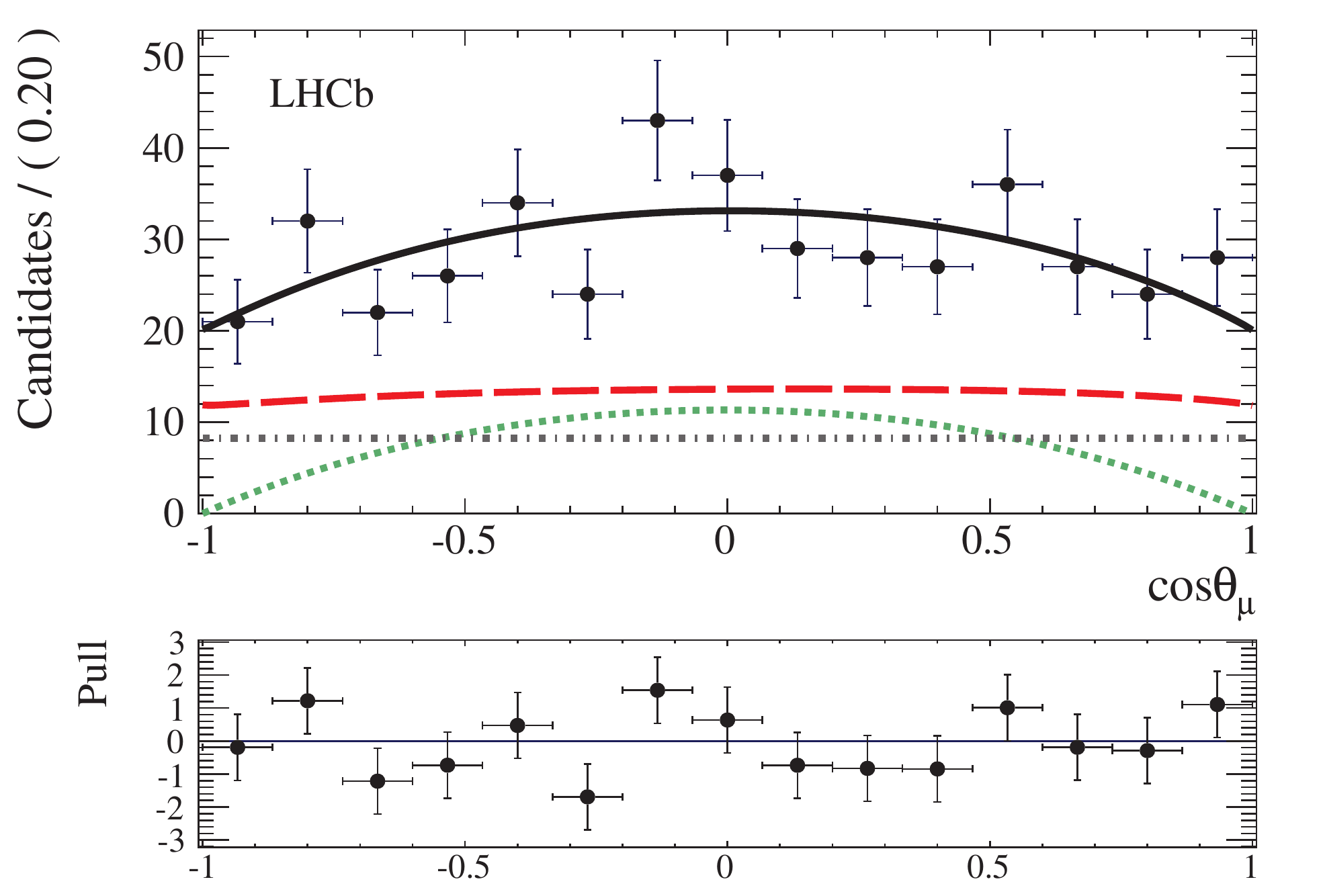} 
  \put(80,59){(a)}
  \end{overpic}
    \begin{overpic}[scale=0.41]{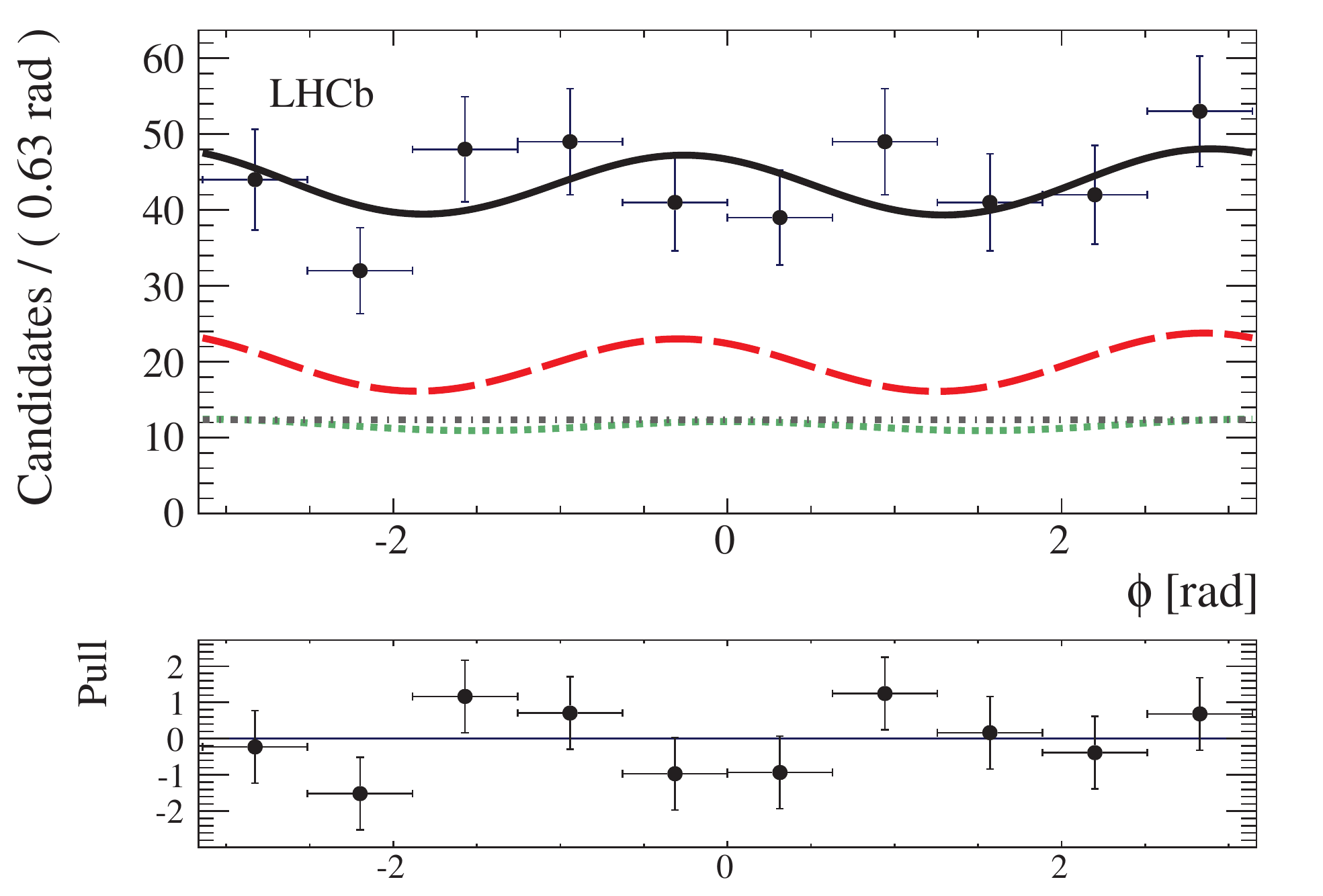} 
  \put(80,59){(b)}
  \end{overpic}
      \begin{overpic}[scale=0.41]{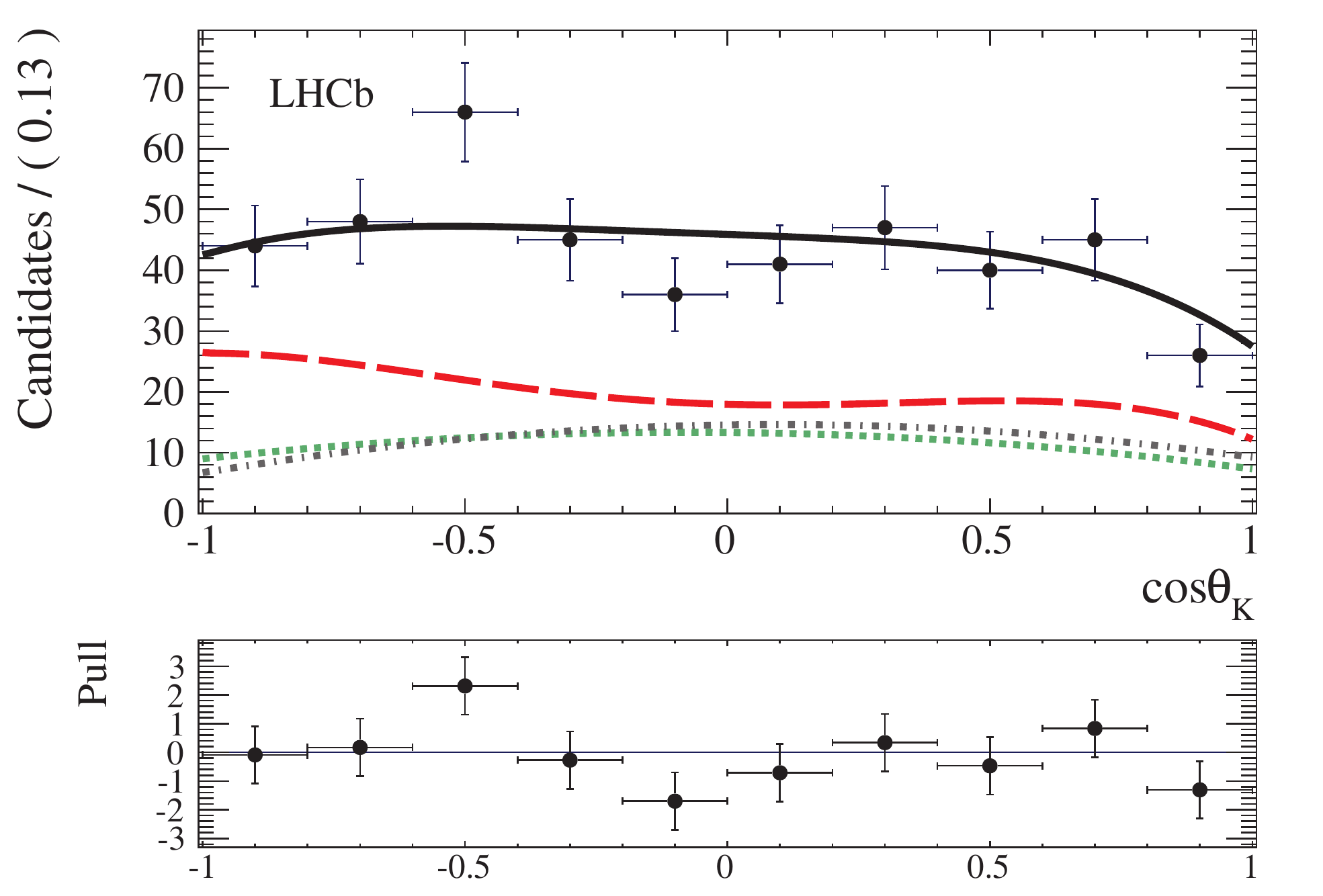} 
  \put(80,59){(c)}
  \end{overpic}
      \begin{overpic}[scale=0.41]{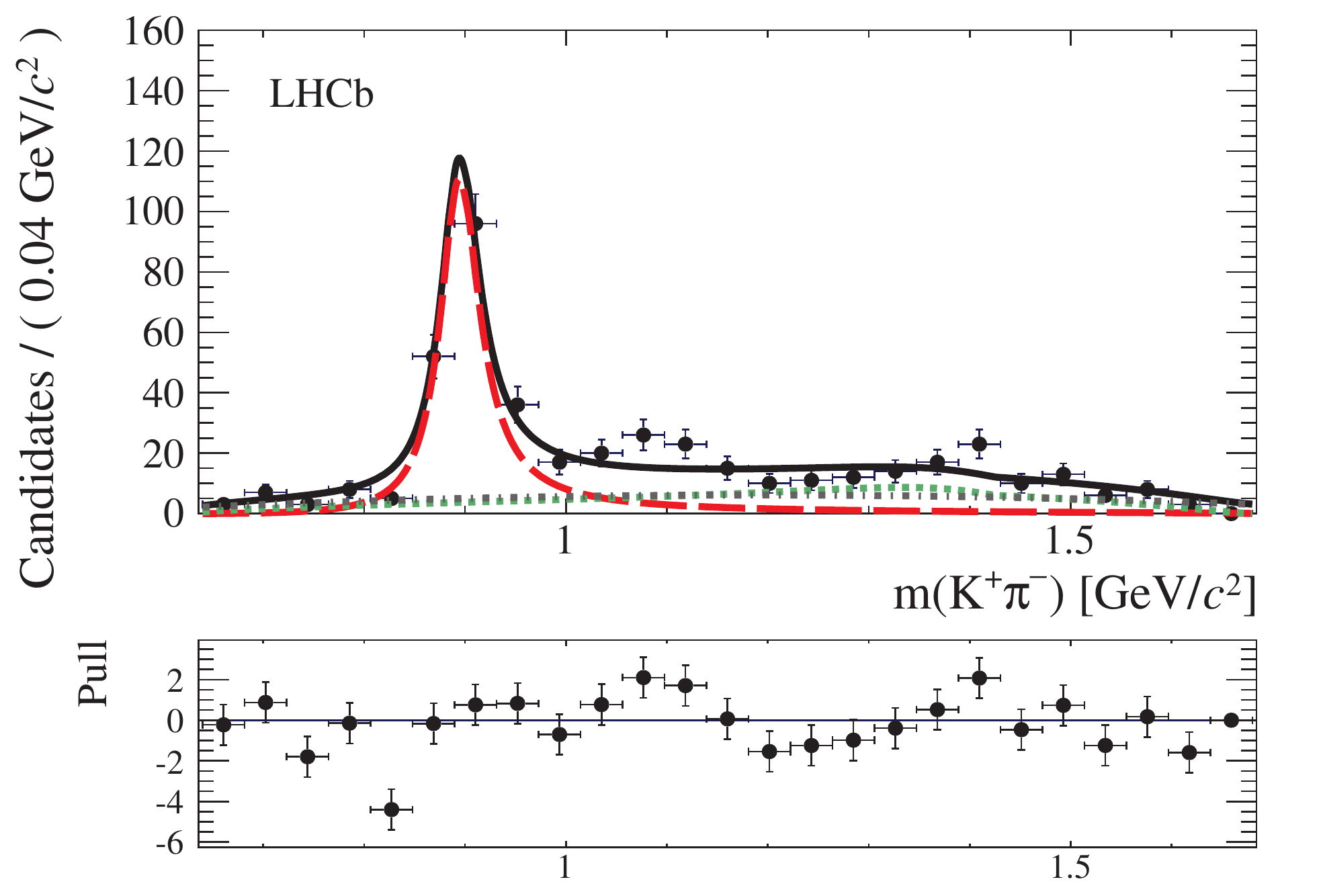} 
  \put(80,59){(d)}
  \end{overpic}
  \caption{\small 
    Distributions of (a) $\cos\theta_\mu$, (b) $\phi$, (c) $\cos\theta_K$ and (d) $m(\Kp\pi^-)$ 
    for selected $\Bsb\to\psi(2S)\Kp\pi^-$ candidates (black points)  with the
  projections of the fitted amplitude model overlaid. The following components are included in the model:
  $K^*(892)^0$  (red dashed), LASS S-wave (green dotted), 
  and background (grey dashed-dotted). The residual pulls are shown below each distribution.
      \label{fig:nominal_fit}
  }
\end{figure}

The default amplitude model is constructed using contributions from
the $\Kstar(892)^0$ resonance and a $\Kp\pi^-$ S-wave
modelled using the LASS parameterisation~\cite{Aston:1987ir}. The 
magnitudes and phases of all components are measured relative to those
of the zero helicity state of the $\Kstar(892)^0$ meson
and the masses and widths of the resonances are fixed to their
known values \cite{PDG2014}.
The remaining eight free parameters are determined using a maximum
likelihood fit of the amplitude to the data in the signal window.
The background fraction is fixed to $0.28$, as determined from the
fit described in Sect.~\ref{sec:massfit}.
The fit fraction for any resonance $R$ is defined in the full 
phase space,
as $f_R = \int S_R\, {\rm d}m_{K\pi}{\rm d}\Omega\, / \int S\, {\rm d}m_{K\pi}{\rm d}\Omega$, where
$S_R$ is the signal amplitude with all amplitude terms set to zero except
those for $R$.
The fractions of each component determined
by the  fit are $f_{K^*(892)^0} = 0.645\pm0.049$, 
and $f_{\rm S-wave} = 0.339\pm0.052$,
where the uncertainty is statistical only. 
The fractions do not sum to unity due to interference between the
different components.
Variations of the S-wave description and 
default mixture of $\Kp\pi^-$ resonances,
including the introduction of the spin-2 $K^*_2(1430)^0$ meson or
an exotic $Z^{-}_{c}$ meson, 
are considered but found to give larger values of the
Poisson likelihood $\chi^2$~\cite{Baker:1983tu}
per degree of freedom or lead to components with fit fractions that
are consistent with zero. For each model the number
of degrees of freedom is calibrated using simulated experiments.
The variations in amplitude model are considered as
sources of systematic uncertainty.
The longitudinal polarisation fraction of the $K^*(892)^0$ meson is defined as
\mbox{$f_{\rm L} = H_0^2/(H_0^2 + H_+^2 + H_-^2)$}, where $H_{0, +, -}$ are the 
magnitudes of the $K^*(892)^0$ helicity amplitudes. This is measured to be
$f_{\rm L} = 0.524 \pm 0.056$, where the uncertainty is statistical.
The projections of the default fit for the helicity angles and invariant
$K^+ \pi^-$ mass are shown in Fig.~\ref{fig:nominal_fit}.

\subsection{Systematic uncertainties of amplitude analysis}
\label{sec:amplitude_syst}

\begin{table}[t]
\caption{\small Summary of systematic uncertainties on the $K^*(892)^0$ fit fraction and $f_{\rm L}$.
Rows marked with (*) refer to uncorrelated sources of uncertainty between the $\Bs$ and $\Bd$ modes
for the computation of the ratio of branching fractions.}
\label{tab:systematics_amp}
\begin{center}
\begin{tabular}{lcc}
Source		&	$K^*(892)^0$ fit fraction	&	$f_{\rm L}$\\
\hline
(*) $\Kp\pi^-$ amplitude model					&		0.028	&	0.017	\\ 
(*) S-wave model								&		0.018	&	0.010	\\ 
$K^*$ resonance widths					&		0.005	&	0.008	\\ 
Blatt-Weisskopf radius						&		0.014	&	0.003	\\ 
Breit-Wigner parameters ($m_R\ {\rm vs.}\ m_{K\pi}$)&		0.026	&	0.005	\\ 
(*) Background parameterisation					&		0.014	&	0.012	\\ 
(*) Background normalisation						&		0.007	&	0.011	\\ 
Efficiency	model (parameterisation)				&		0.011	&	0.007	\\ 
Efficiency	model (neural net)					&		0.002	&	0.004	\\ 
\hline
Quadrature sum of systematic uncertainties			&		0.049	&	0.029	\\
Quadrature sum of uncorrelated systematic uncertainties	&		0.037	&	0.026	\\
\hline
Statistical uncertainty					&		0.049	&	0.056	\\
\hline
\end{tabular}
\end{center}
\end{table}

A summary of possible sources of systematic uncertainties that affect the amplitude 
analysis is reported in Table~\ref{tab:systematics_amp}. The size of each 
contribution is determined using a set of simulated experiments, of the same size as the data,
generated
under the hypothesis of an alternative amplitude model. These are fitted once with the default model and 
again with the alternative model. The experiment-by-experiment difference in the measured
fit fractions and $f_{\rm L}$ is then computed 
and the sum in quadrature of the mean and standard deviation is assigned as a systematic uncertainty to the
corresponding parameter.

The systematic dependence on the $\Kp\pi^-$ amplitude model is determined
using the above procedure, where the alternative model also contains a spin-2 $K^*_2(1430)$
component. This leads to the dominant systematic uncertainty on the $K^*(892)^0$
fit fraction and $f_{\rm L}$.
The systematic dependence on the $\Kp\pi^-$ S-wave model is determined
using simulated experiments where a combination of a non-resonant term and
a $K^*_0(1430)$ contribution is used in place of the LASS parameterisation. 
In addition, the amplitude model contains parameters that are fixed in the default fit such
as the masses and widths of the resonances and the Blatt-Weisskopf radius. The radius
controls the effective hadron size and is set to $1.6 \,$(GeV$/c$)$^{-1}$ by default. Alternative
models are considered where this is changed to $3.0 \,$(GeV$/c$)$^{-1}$ and
$0.8 \,$(GeV$/c$)$^{-1}$.

A large source of systematic uncertainty comes from the choice of convention for
the mass, $m$, in the
$(p/m)^{L_R}$ terms of the amplitude. The default amplitude model follows 
the convention in Ref.~\cite{PDG2014} by using the resonance mass. This
is different to that in Ref.~\cite{LHCb-PAPER-2014-014}
where the running resonance mass ($m_{K\pi}$) is used in the 
denominator. This choice is motivated by the improved fit quality obtained when
using the resonance mass.

The systematic uncertainty related to the combinatorial background
parameterisation is determined using an amplitude model with an
alternative background description that allows for higher moment
contributions ($a\le2$, $b\le2$, $c\le2$ and $d\le2$).
The combinatorial background normalisation is determined from the
fit to the $m(\psi(2S)K^+\pi^-)$ distribution and is fixed in the amplitude fit. The systematic
uncertainty related to the level of the background is estimated by using an
amplitude model with the background fraction modified by $\pm10\%$.

The efficiency parameterisation is tested by re-evaluating the coefficients, allowing
for higher order moments ($a\le4$, $b\le4$, $c\le4$ and $d\le4$). Similarly, to test the
dependence of the efficiency model on the neural network requirement, an alternative model is used
with the efficiency parameterisation determined from the simulated
events that are selected without applying the requirement.
There is a negligible systematic uncertainty caused by the lifetime difference between the \Bd and \Bs mesons.

\section{Branching fraction results}
\label{sec:systematics}
Two ratios of branching fractions are calculated, ${\cal B}(\overline{B}^0_s \rightarrow \psi(2S) K^+ \pi^-)/{\cal B}(B^0
  \rightarrow \psi(2S) K^+ \pi^-)$ and ${\cal B}(\overline{B}^0_s \rightarrow \psi(2S) K^*(892)^0)/{\cal B}(B^0
  \rightarrow \psi(2S) K^*(892)^0)$. These are determined from the
signal yields given in Sect.~\ref{sec:massfit} correcting for the
relative detector acceptance using simulation. The
simulated $\Bs$ samples are reweighted with the results of the
angular analysis presented in Sect.~\ref{sec:angular}. Similarly,
the $\Bz$ simulated data are reweighted to match the results given in 
Ref. \cite{LHCb-PAPER-2014-014}. For the inclusive branching ratio, the
relative efficiency between the two modes is found to be $0.975
\pm 0.014$ whilst for the $K^*(892)^0$ component it is $1.027 \pm
0.021$. The uncertainty on these values is propagated to the
systematic uncertainty.

Since the same final state is considered in the signal and
normalisation mode, most sources of systematic 
uncertainty cancel in the ratio. The remaining
sources are discussed in the following.
 The variations of the invariant mass fit model described in
 Sect.~\ref{sec:massfit} are considered. The largest change in the ratio
of yields observed in these tests is $3.7\%$, which is assigned as a
systematic uncertainty. Differences in the $\pt$ spectra of the \Bd
meson are seen comparing data and
the reweighted simulation. If the $\pt$ spectrum in the simulation is
further reweighted to match the data, the efficiency ratio changes by $0.7 \%$,
which is assigned as a systematic uncertainty.
 
To test the impact of the chosen $\Kp\pi^-$ amplitude model for the $\Bs$ channel,
the simulated events are reweighted using a model consisting of the $K^*(892)^0$
resonance, the LASS~\cite{Aston:1987ir} description of the S-wave and the $K^*_2(1430)$ resonance.
This changes
the efficiency ratio by $0.6 \%$, which is assigned as a systematic
uncertainty. To calculate the $\Kstar(892)^0$ branching ratio, the fraction of
candidates from this source is needed. For the $\Bs$ channel this is
determined from the amplitude analysis to be $0.645 \pm 0.049 \pm 0.049$
and the corresponding fraction for the $\Bz$ channel is $0.591 \pm
0.009$ \cite{LHCb-PAPER-2014-014}, leading to a $6.0\%$ systematic
uncertainty. All of the uncertainties discussed above are
summarised in Table~\ref{tab:systratio}.
The limited knowledge of the fragmentation fractions, $f_s/f_d   = 0.259 \pm 0.015$
\cite{LHCb-PAPER-2011-018,LHCb-PAPER-2012-037, LHCb-CONF-2013-011},
results in an uncertainty of $5.8\%$, which is quoted separately from
the others.
\begin{table}[t]
\caption{\small Systematic uncertainties on the ratio of branching fractions.}
\label{tab:systratio}
\begin{center}
\small
\begin{tabular}{lc|c} 
 			&  \multicolumn{2}{c}{Relative uncertainty \%} \\
Source & Inclusive & $\Kstar(892)^0$  \\
\hline
Simulation sample size    & 1.4  & 2.2 \\
Fit model                  &  3.7 & 3.7 \\
Detector acceptance         & 0.7 & 0.7 \\
$\Kp\pi^-$ amplitude model & 0.6 &  --\\
$\Kstar(892)^0$ fit fraction &  -- & 6.0 \\
\hline
Quadrature sum                          & 4.1 & 7.4 \\
\hline
\end{tabular}
\end{center}
\end{table}
%

\section{Summary}
\label{sec:conclusions}
Using a data set corresponding to an integrated luminosity of $3.0 \invfb$ collected in $pp$ collisions at
centre-of-mass energies of 7 and 8 TeV, the decay $\overline{B}_s^0 \rightarrow \psi(2S) K^+ \pi^-$ is
observed.
The mass splitting between the $\Bs$ and $\Bz$ mesons is 
measured to be
\begin{equation}
M(\Bsb) -M(\Bz)  = 87.45 \pm 0.44 \stat \pm 0.09 \syst
\mevcc. \nonumber
\end{equation}
This is consistent with, though less precise than, the value $87.21
\pm 0.31 \mevcc$  obtained by  averaging the
results in Refs.~\cite{LHCb-PAPER-2011-035, CDFMasses}. Averaging the
two numbers gives 
\begin{equation}
M(\Bsb) -M(\Bz)  = 87.29 \pm 0.26 \mevcc. \nonumber
\end{equation}
The ratio of branching fractions
between the $\Bsb$ and $\Bz$ modes is measured to be
\begin{equation}
\frac{{\cal B}(\overline{B}^0_s \rightarrow \psi(2S) K^+ \pi^-)}{{\cal B}(B^0
  \rightarrow \psi(2S) K^+ \pi^-)} = 5.38 \pm  0.36 \stat \pm 0.22 \syst
\pm 0.31  \, (f_s/f_d ) \, \%.\nonumber
\end{equation}
The fraction of decays proceeding via an intermediate $K^*(892)^0$ meson
is measured with an amplitude analysis to be $0.645 \pm 0.049 \stat \pm 0.049 \syst$.
No significant structure is seen in the distribution of $m(\psi(2S)\pi^-)$.

The longitudinal polarisation fraction, $f_{\rm L}$, of the $K^*(892)^0$ meson
is determined as $0.524 \pm 0.056 \stat \pm 0.029 \syst$. 
This is consistent with the value measured in the corresponding
decay that proceeds through an intermediate $J/\psi$ meson,
\mbox{$f_{\rm L} = 0.50 \pm 0.08 \stat \pm 0.02 \syst$}
\cite{LHCb-PAPER-2012-014}. The present data set does
not allow a test of the prediction given in Ref.~\cite{Hiller:2013cza} that
$f_{\rm L}$ should be lower for decays closer to the kinematic endpoint. 

Using the $K^*(892)^0$ fraction 
determined in this analysis for the $\Bs$ component, the
corresponding number for the $\Bz$ mode from Ref.~\cite{LHCb-PAPER-2014-014},
and the efficiency ratio given in Sect.~\ref{sec:systematics},
the following ratio of branching fractions is measured
\begin{equation}
\frac{{\cal B}(\overline{B}^0_s \rightarrow \psi(2S) K^*(892)^0)}{{\cal B}(B^0
  \rightarrow \psi(2S) K^*(892)^0)} = 5.58 \pm  0.57 \stat \pm 0.40 \syst
\pm 0.32  \, (f_s/f_d)  \, \%.\nonumber
\end{equation}
The $\overline{B}^0_s \rightarrow \psi(2S) K^+ \pi^-$ mode may
be useful for future studies
that attempt to control the size of loop-mediated processes 
that influence \CP violation studies and offers promising
opportunities in the search for exotic resonances.

\section*{Acknowledgements}
 
\noindent We express our gratitude to our colleagues in the CERN
accelerator departments for the excellent performance of the LHC. We
thank the technical and administrative staff at the LHCb
institutes. We acknowledge support from CERN and from the national
agencies: CAPES, CNPq, FAPERJ and FINEP (Brazil); NSFC (China);
CNRS/IN2P3 (France); BMBF, DFG, HGF and MPG (Germany); INFN (Italy); 
FOM and NWO (The Netherlands); MNiSW and NCN (Poland); MEN/IFA (Romania); 
MinES and FANO (Russia); MinECo (Spain); SNSF and SER (Switzerland); 
NASU (Ukraine); STFC (United Kingdom); NSF (USA).
The Tier1 computing centres are supported by IN2P3 (France), KIT and BMBF 
(Germany), INFN (Italy), NWO and SURF (The Netherlands), PIC (Spain), GridPP 
(United Kingdom).
We are indebted to the communities behind the multiple open 
source software packages on which we depend. We are also thankful for the 
computing resources and the access to software R\&D tools provided by Yandex LLC (Russia).
Individual groups or members have received support from 
EPLANET, Marie Sk\l{}odowska-Curie Actions and ERC (European Union), 
Conseil g\'{e}n\'{e}ral de Haute-Savoie, Labex ENIGMASS and OCEVU, 
R\'{e}gion Auvergne (France), RFBR (Russia), XuntaGal and GENCAT (Spain), Royal Society and Royal
Commission for the Exhibition of 1851 (United Kingdom).

\addcontentsline{toc}{section}{References}
\setboolean{inbibliography}{true}
\bibliographystyle{LHCb}

\begin{mcitethebibliography}{10}
\mciteSetBstSublistMode{n}
\mciteSetBstMaxWidthForm{subitem}{\alph{mcitesubitemcount})}
\mciteSetBstSublistLabelBeginEnd{\mcitemaxwidthsubitemform\space}
{\relax}{\relax}

\bibitem{LHCb-PAPER-2014-059}
LHCb collaboration, R.~Aaij {\em et~al.},
  \ifthenelse{\boolean{articletitles}}{{\it {Precision measurement of $CP$
  violation in $B_s^0 \to J/\psi K^+K^-$ decays}},
  }{}\href{http://dx.doi.org/10.1103/PhysRevLett.114.041801}{Phys.\ Rev.\
  Lett.\  {\bf 114} (2015) 041801}, \href{http://arxiv.org/abs/1411.3104}{{\tt
  arXiv:1411.3104}}\relax
\mciteBstWouldAddEndPuncttrue
\mciteSetBstMidEndSepPunct{\mcitedefaultmidpunct}
{\mcitedefaultendpunct}{\mcitedefaultseppunct}\relax
\EndOfBibitem
\bibitem{LHCb-PAPER-2012-006}
LHCb collaboration, R.~Aaij {\em et~al.},
  \ifthenelse{\boolean{articletitles}}{{\it {Measurement of the $CP$-violating
  phase $\phi_s$ in $\bar{B}^0_s \to J/\psi \pi^+\pi^-$ decays}},
  }{}\href{http://dx.doi.org/10.1016/j.physletb.2012.06.032}{Phys.\ Lett.\
  {\bf B713} (2012) 378}, \href{http://arxiv.org/abs/1204.5675}{{\tt
  arXiv:1204.5675}}\relax
\mciteBstWouldAddEndPuncttrue
\mciteSetBstMidEndSepPunct{\mcitedefaultmidpunct}
{\mcitedefaultendpunct}{\mcitedefaultseppunct}\relax
\EndOfBibitem
\bibitem{Faller:2008gt}
S.~Faller, R.~Fleischer, and T.~Mannel,
  \ifthenelse{\boolean{articletitles}}{{\it {Precision physics with $B^0_s \to
  J/\psi \phi$ at the LHC: the quest for new physics}},
  }{}\href{http://dx.doi.org/10.1103/PhysRevD.79.014005}{Phys.\ Rev.\  {\bf
  D79} (2009) 014005}, \href{http://arxiv.org/abs/0810.4248}{{\tt
  arXiv:0810.4248}}\relax
\mciteBstWouldAddEndPuncttrue
\mciteSetBstMidEndSepPunct{\mcitedefaultmidpunct}
{\mcitedefaultendpunct}{\mcitedefaultseppunct}\relax
\EndOfBibitem
\bibitem{Aaltonen:2011sy}
CDF collaboration, T.~Aaltonen {\em et~al.},
  \ifthenelse{\boolean{articletitles}}{{\it {Observation of $B^0_s \to J/\psi
  K^{*0}(892)$ and $B^0_s \to J/\psi K^0_S$ decays}},
  }{}\href{http://dx.doi.org/10.1103/PhysRevD.83.052012}{Phys.\ Rev.\  {\bf
  D83} (2011) 052012}, \href{http://arxiv.org/abs/1102.1961}{{\tt
  arXiv:1102.1961}}\relax
\mciteBstWouldAddEndPuncttrue
\mciteSetBstMidEndSepPunct{\mcitedefaultmidpunct}
{\mcitedefaultendpunct}{\mcitedefaultseppunct}\relax
\EndOfBibitem
\bibitem{LHCb-PAPER-2012-014}
LHCb collaboration, R.~Aaij {\em et~al.},
  \ifthenelse{\boolean{articletitles}}{{\it {Measurement of the $B^0_s \to
  J/\psi \bar{K}^{*0}$ branching fraction and angular amplitudes}},
  }{}\href{http://dx.doi.org/10.1103/PhysRevD.86.071102}{Phys.\ Rev.\  {\bf
  D86} (2012) 071102(R)}, \href{http://arxiv.org/abs/1208.0738}{{\tt
  arXiv:1208.0738}}\relax
\mciteBstWouldAddEndPuncttrue
\mciteSetBstMidEndSepPunct{\mcitedefaultmidpunct}
{\mcitedefaultendpunct}{\mcitedefaultseppunct}\relax
\EndOfBibitem
\bibitem{Choi:2007wga}
Belle collaboration, S.~Choi {\em et~al.},
  \ifthenelse{\boolean{articletitles}}{{\it {Observation of a resonance-like
  structure in the $\pi^{\pm} \psi'$ mass distribution in exclusive $B\to K
  \pi^{\pm} \psi'$ decays}},
  }{}\href{http://dx.doi.org/10.1103/PhysRevLett.100.142001}{Phys.\ Rev.\
  Lett.\  {\bf 100} (2008) 142001}, \href{http://arxiv.org/abs/0708.1790}{{\tt
  arXiv:0708.1790}}\relax
\mciteBstWouldAddEndPuncttrue
\mciteSetBstMidEndSepPunct{\mcitedefaultmidpunct}
{\mcitedefaultendpunct}{\mcitedefaultseppunct}\relax
\EndOfBibitem
\bibitem{Mizuk:2009da}
Belle collaboration, R.~Mizuk {\em et~al.},
  \ifthenelse{\boolean{articletitles}}{{\it {Dalitz analysis of $B\to K \pi^+
  \psi'$ decays and the $Z(4430)^+$}},
  }{}\href{http://dx.doi.org/10.1103/PhysRevD.80.031104}{Phys.\ Rev.\  {\bf
  D80} (2009) 031104}, \href{http://arxiv.org/abs/0905.2869}{{\tt
  arXiv:0905.2869}}\relax
\mciteBstWouldAddEndPuncttrue
\mciteSetBstMidEndSepPunct{\mcitedefaultmidpunct}
{\mcitedefaultendpunct}{\mcitedefaultseppunct}\relax
\EndOfBibitem
\bibitem{Chilikin:2013tch}
Belle collaboration, K.~Chilikin {\em et~al.},
  \ifthenelse{\boolean{articletitles}}{{\it {Experimental constraints on the
  spin and parity of the $Z(4430)^+$}},
  }{}\href{http://dx.doi.org/10.1103/PhysRevD.88.074026}{Phys.\ Rev.\  {\bf
  D88} (2013) 074026}, \href{http://arxiv.org/abs/1306.4894}{{\tt
  arXiv:1306.4894}}\relax
\mciteBstWouldAddEndPuncttrue
\mciteSetBstMidEndSepPunct{\mcitedefaultmidpunct}
{\mcitedefaultendpunct}{\mcitedefaultseppunct}\relax
\EndOfBibitem
\bibitem{LHCb-PAPER-2014-014}
LHCb collaboration, R.~Aaij {\em et~al.},
  \ifthenelse{\boolean{articletitles}}{{\it {Observation of the resonant
  character of the $Z(4430)^-$ state}},
  }{}\href{http://dx.doi.org/10.1103/PhysRevLett.112.222002}{Phys.\ Rev.\
  Lett.\  {\bf 112} (2014) 222002}, \href{http://arxiv.org/abs/1404.1903}{{\tt
  arXiv:1404.1903}}\relax
\mciteBstWouldAddEndPuncttrue
\mciteSetBstMidEndSepPunct{\mcitedefaultmidpunct}
{\mcitedefaultendpunct}{\mcitedefaultseppunct}\relax
\EndOfBibitem
\bibitem{Klempt:2007cp}
E.~Klempt and A.~Zaitsev, \ifthenelse{\boolean{articletitles}}{{\it {Glueballs,
  hybrids, multiquarks. Experimental facts versus QCD inspired concepts}},
  }{}\href{http://dx.doi.org/10.1016/j.physrep.2007.07.006}{Phys.\ Rept.\  {\bf
  454} (2007) 1}, \href{http://arxiv.org/abs/0708.4016}{{\tt
  arXiv:0708.4016}}\relax
\mciteBstWouldAddEndPuncttrue
\mciteSetBstMidEndSepPunct{\mcitedefaultmidpunct}
{\mcitedefaultendpunct}{\mcitedefaultseppunct}\relax
\EndOfBibitem
\bibitem{Mizuk:2008me}
Belle collaboration, R.~Mizuk {\em et~al.},
  \ifthenelse{\boolean{articletitles}}{{\it {Observation of two resonance-like
  structures in the $\pi^+ \chi_{c1}$ mass distribution in exclusive
  $\bar{B}^0\to K^- \pi^+ \chi_{c1}$ decays}},
  }{}\href{http://dx.doi.org/10.1103/PhysRevD.78.072004}{Phys.\ Rev.\  {\bf
  D78} (2008) 072004}, \href{http://arxiv.org/abs/0806.4098}{{\tt
  arXiv:0806.4098}}\relax
\mciteBstWouldAddEndPuncttrue
\mciteSetBstMidEndSepPunct{\mcitedefaultmidpunct}
{\mcitedefaultendpunct}{\mcitedefaultseppunct}\relax
\EndOfBibitem
\bibitem{Chilikin:2014bkk}
Belle collaboration, K.~Chilikin {\em et~al.},
  \ifthenelse{\boolean{articletitles}}{{\it {Observation of a new charged
  charmonium-like state in $B^0 \to J/\psi K^-\pi^+$ decays}},
  }{}\href{http://dx.doi.org/10.1103/PhysRevD.90.112009}{Phys.\ Rev.\  {\bf
  D90} (2014) 112009}, \href{http://arxiv.org/abs/1408.6457}{{\tt
  arXiv:1408.6457}}\relax
\mciteBstWouldAddEndPuncttrue
\mciteSetBstMidEndSepPunct{\mcitedefaultmidpunct}
{\mcitedefaultendpunct}{\mcitedefaultseppunct}\relax
\EndOfBibitem
\bibitem{Alves:2008zz}
LHCb collaboration, A.~A. Alves~Jr. {\em et~al.},
  \ifthenelse{\boolean{articletitles}}{{\it {The \lhcb detector at the LHC}},
  }{}\href{http://dx.doi.org/10.1088/1748-0221/3/08/S08005}{JINST {\bf 3}
  (2008) S08005}\relax
\mciteBstWouldAddEndPuncttrue
\mciteSetBstMidEndSepPunct{\mcitedefaultmidpunct}
{\mcitedefaultendpunct}{\mcitedefaultseppunct}\relax
\EndOfBibitem
\bibitem{LHCb-DP-2014-002}
LHCb collaboration, R.~Aaij {\em et~al.},
  \ifthenelse{\boolean{articletitles}}{{\it {LHCb detector performance}},
  }{}\href{http://dx.doi.org/10.1142/S0217751X15300227}{Int.\ J.\ Mod.\ Phys.\
  {\bf A30} (2015) 1530022}, \href{http://arxiv.org/abs/1412.6352}{{\tt
  arXiv:1412.6352}}\relax
\mciteBstWouldAddEndPuncttrue
\mciteSetBstMidEndSepPunct{\mcitedefaultmidpunct}
{\mcitedefaultendpunct}{\mcitedefaultseppunct}\relax
\EndOfBibitem
\bibitem{LHCb-DP-2013-003}
R.~Arink {\em et~al.}, \ifthenelse{\boolean{articletitles}}{{\it {Performance
  of the LHCb Outer Tracker}},
  }{}\href{http://dx.doi.org/10.1088/1748-0221/9/01/P01002}{JINST {\bf 9}
  (2014) P01002}, \href{http://arxiv.org/abs/1311.3893}{{\tt
  arXiv:1311.3893}}\relax
\mciteBstWouldAddEndPuncttrue
\mciteSetBstMidEndSepPunct{\mcitedefaultmidpunct}
{\mcitedefaultendpunct}{\mcitedefaultseppunct}\relax
\EndOfBibitem
\bibitem{LHCb-PAPER-2012-048}
LHCb collaboration, R.~Aaij {\em et~al.},
  \ifthenelse{\boolean{articletitles}}{{\it {Measurements of the $\Lambda_b^0$,
  $\Xi_b^-$, and $\Omega_b^-$ baryon masses}},
  }{}\href{http://dx.doi.org/10.1103/PhysRevLett.110.182001}{Phys.\ Rev.\
  Lett.\  {\bf 110} (2013) 182001}, \href{http://arxiv.org/abs/1302.1072}{{\tt
  arXiv:1302.1072}}\relax
\mciteBstWouldAddEndPuncttrue
\mciteSetBstMidEndSepPunct{\mcitedefaultmidpunct}
{\mcitedefaultendpunct}{\mcitedefaultseppunct}\relax
\EndOfBibitem
\bibitem{LHCb-DP-2012-003}
M.~Adinolfi {\em et~al.}, \ifthenelse{\boolean{articletitles}}{{\it
  {Performance of the \lhcb RICH detector at the LHC}},
  }{}\href{http://dx.doi.org/10.1140/epjc/s10052-013-2431-9}{Eur.\ Phys.\ J.\
  {\bf C73} (2013) 2431}, \href{http://arxiv.org/abs/1211.6759}{{\tt
  arXiv:1211.6759}}\relax
\mciteBstWouldAddEndPuncttrue
\mciteSetBstMidEndSepPunct{\mcitedefaultmidpunct}
{\mcitedefaultendpunct}{\mcitedefaultseppunct}\relax
\EndOfBibitem
\bibitem{LHCb-DP-2012-002}
A.~A. Alves~Jr. {\em et~al.}, \ifthenelse{\boolean{articletitles}}{{\it
  {Performance of the LHCb muon system}},
  }{}\href{http://dx.doi.org/10.1088/1748-0221/8/02/P02022}{JINST {\bf 8}
  (2013) P02022}, \href{http://arxiv.org/abs/1211.1346}{{\tt
  arXiv:1211.1346}}\relax
\mciteBstWouldAddEndPuncttrue
\mciteSetBstMidEndSepPunct{\mcitedefaultmidpunct}
{\mcitedefaultendpunct}{\mcitedefaultseppunct}\relax
\EndOfBibitem
\bibitem{LHCb-DP-2012-004}
R.~Aaij {\em et~al.}, \ifthenelse{\boolean{articletitles}}{{\it {The \lhcb
  trigger and its performance in 2011}},
  }{}\href{http://dx.doi.org/10.1088/1748-0221/8/04/P04022}{JINST {\bf 8}
  (2013) P04022}, \href{http://arxiv.org/abs/1211.3055}{{\tt
  arXiv:1211.3055}}\relax
\mciteBstWouldAddEndPuncttrue
\mciteSetBstMidEndSepPunct{\mcitedefaultmidpunct}
{\mcitedefaultendpunct}{\mcitedefaultseppunct}\relax
\EndOfBibitem
\bibitem{Sjostrand:2006za}
T.~Sj\"{o}strand, S.~Mrenna, and P.~Skands,
  \ifthenelse{\boolean{articletitles}}{{\it {\pythia 6.4 physics and manual}},
  }{}\href{http://dx.doi.org/10.1088/1126-6708/2006/05/026}{JHEP {\bf 05}
  (2006) 026}, \href{http://arxiv.org/abs/hep-ph/0603175}{{\tt
  arXiv:hep-ph/0603175}}\relax
\mciteBstWouldAddEndPuncttrue
\mciteSetBstMidEndSepPunct{\mcitedefaultmidpunct}
{\mcitedefaultendpunct}{\mcitedefaultseppunct}\relax
\EndOfBibitem
\bibitem{Sjostrand:2007gs}
T.~Sj\"{o}strand, S.~Mrenna, and P.~Skands,
  \ifthenelse{\boolean{articletitles}}{{\it {A brief introduction to \pythia
  8.1}}, }{}\href{http://dx.doi.org/10.1016/j.cpc.2008.01.036}{Comput.\ Phys.\
  Commun.\  {\bf 178} (2008) 852}, \href{http://arxiv.org/abs/0710.3820}{{\tt
  arXiv:0710.3820}}\relax
\mciteBstWouldAddEndPuncttrue
\mciteSetBstMidEndSepPunct{\mcitedefaultmidpunct}
{\mcitedefaultendpunct}{\mcitedefaultseppunct}\relax
\EndOfBibitem
\bibitem{LHCb-PROC-2010-056}
I.~Belyaev {\em et~al.}, \ifthenelse{\boolean{articletitles}}{{\it {Handling of
  the generation of primary events in \gauss, the LHCb simulation framework}},
  }{}\href{http://dx.doi.org/10.1109/NSSMIC.2010.5873949}{Nuclear Science
  Symposium Conference Record (NSS/MIC) {\bf IEEE} (2010) 1155}\relax
\mciteBstWouldAddEndPuncttrue
\mciteSetBstMidEndSepPunct{\mcitedefaultmidpunct}
{\mcitedefaultendpunct}{\mcitedefaultseppunct}\relax
\EndOfBibitem
\bibitem{Lange:2001uf}
D.~J. Lange, \ifthenelse{\boolean{articletitles}}{{\it {The \evtgen particle
  decay simulation package}},
  }{}\href{http://dx.doi.org/10.1016/S0168-9002(01)00089-4}{Nucl.\ Instrum.\
  Meth.\  {\bf A462} (2001) 152}\relax
\mciteBstWouldAddEndPuncttrue
\mciteSetBstMidEndSepPunct{\mcitedefaultmidpunct}
{\mcitedefaultendpunct}{\mcitedefaultseppunct}\relax
\EndOfBibitem
\bibitem{Golonka:2005pn}
P.~Golonka and Z.~Was, \ifthenelse{\boolean{articletitles}}{{\it {\photos Monte
  Carlo: A precision tool for QED corrections in $Z$ and $W$ decays}},
  }{}\href{http://dx.doi.org/10.1140/epjc/s2005-02396-4}{Eur.\ Phys.\ J.\  {\bf
  C45} (2006) 97}, \href{http://arxiv.org/abs/hep-ph/0506026}{{\tt
  arXiv:hep-ph/0506026}}\relax
\mciteBstWouldAddEndPuncttrue
\mciteSetBstMidEndSepPunct{\mcitedefaultmidpunct}
{\mcitedefaultendpunct}{\mcitedefaultseppunct}\relax
\EndOfBibitem
\bibitem{Allison:2006ve}
\geant collaboration, J.~Allison {\em et~al.},
  \ifthenelse{\boolean{articletitles}}{{\it {\geant developments and
  applications}}, }{}\href{http://dx.doi.org/10.1109/TNS.2006.869826}{IEEE
  Trans.\ Nucl.\ Sci.\  {\bf 53} (2006) 270}\relax
\mciteBstWouldAddEndPuncttrue
\mciteSetBstMidEndSepPunct{\mcitedefaultmidpunct}
{\mcitedefaultendpunct}{\mcitedefaultseppunct}\relax
\EndOfBibitem
\bibitem{Agostinelli:2002hh}
\geant collaboration, S.~Agostinelli {\em et~al.},
  \ifthenelse{\boolean{articletitles}}{{\it {\geant: a simulation toolkit}},
  }{}\href{http://dx.doi.org/10.1016/S0168-9002(03)01368-8}{Nucl.\ Instrum.\
  Meth.\  {\bf A506} (2003) 250}\relax
\mciteBstWouldAddEndPuncttrue
\mciteSetBstMidEndSepPunct{\mcitedefaultmidpunct}
{\mcitedefaultendpunct}{\mcitedefaultseppunct}\relax
\EndOfBibitem
\bibitem{LHCb-PROC-2011-006}
M.~Clemencic {\em et~al.}, \ifthenelse{\boolean{articletitles}}{{\it {The \lhcb
  simulation application, \gauss: design, evolution and experience}},
  }{}\href{http://dx.doi.org/10.1088/1742-6596/331/3/032023}{{J.\ Phys.\ Conf.\
  Ser.\ } {\bf 331} (2011) 032023}\relax
\mciteBstWouldAddEndPuncttrue
\mciteSetBstMidEndSepPunct{\mcitedefaultmidpunct}
{\mcitedefaultendpunct}{\mcitedefaultseppunct}\relax
\EndOfBibitem
\bibitem{PDG2014}
Particle Data Group, K.~A. Olive {\em et~al.},
  \ifthenelse{\boolean{articletitles}}{{\it {\href{http://pdg.lbl.gov/}{Review
  of particle physics}}},
  }{}\href{http://dx.doi.org/10.1088/1674-1137/38/9/090001}{Chin.\ Phys.\  {\bf
  C38} (2014) 090001}\relax
\mciteBstWouldAddEndPuncttrue
\mciteSetBstMidEndSepPunct{\mcitedefaultmidpunct}
{\mcitedefaultendpunct}{\mcitedefaultseppunct}\relax
\EndOfBibitem
\bibitem{Hulsbergen:2005pu}
W.~D. Hulsbergen, \ifthenelse{\boolean{articletitles}}{{\it {Decay chain
  fitting with a Kalman filter}},
  }{}\href{http://dx.doi.org/10.1016/j.nima.2005.06.078}{Nucl.\ Instrum.\
  Meth.\  {\bf A552} (2005) 566},
  \href{http://arxiv.org/abs/physics/0503191}{{\tt
  arXiv:physics/0503191}}\relax
\mciteBstWouldAddEndPuncttrue
\mciteSetBstMidEndSepPunct{\mcitedefaultmidpunct}
{\mcitedefaultendpunct}{\mcitedefaultseppunct}\relax
\EndOfBibitem
\bibitem{Skwarnicki:1986xj}
T.~Skwarnicki, {\em {A study of the radiative cascade transitions between the
  Upsilon-prime and Upsilon resonances}}, PhD thesis, Institute of Nuclear
  Physics, Krakow, 1986,
  {\href{http://inspirehep.net/record/230779/files/230779.pdf}{DESY-F31-86-02}}\relax
\mciteBstWouldAddEndPuncttrue
\mciteSetBstMidEndSepPunct{\mcitedefaultmidpunct}
{\mcitedefaultendpunct}{\mcitedefaultseppunct}\relax
\EndOfBibitem
\bibitem{Aston:1987ir}
D.~Aston {\em et~al.}, \ifthenelse{\boolean{articletitles}}{{\it {A study of
  $K^- \pi^+$ scattering in the reaction $K^- p \to K^- \pi^+ n$ at 11 GeV/c}},
  }{}\href{http://dx.doi.org/10.1016/0550-3213(88)90028-4}{Nucl.\ Phys.\  {\bf
  B296} (1988) 493}\relax
\mciteBstWouldAddEndPuncttrue
\mciteSetBstMidEndSepPunct{\mcitedefaultmidpunct}
{\mcitedefaultendpunct}{\mcitedefaultseppunct}\relax
\EndOfBibitem
\bibitem{Baker:1983tu}
S.~Baker and R.~D. Cousins, \ifthenelse{\boolean{articletitles}}{{\it
  Clarification of the use of chi square and likelihood functions in fits to
  histograms}, }{}\href{http://dx.doi.org/10.1016/0167-5087(84)90016-4}{Nucl.\
  Instrum.\ Meth.\  {\bf 221} (1984) 437}\relax
\mciteBstWouldAddEndPuncttrue
\mciteSetBstMidEndSepPunct{\mcitedefaultmidpunct}
{\mcitedefaultendpunct}{\mcitedefaultseppunct}\relax
\EndOfBibitem
\bibitem{LHCb-PAPER-2011-018}
LHCb collaboration, R.~Aaij {\em et~al.},
  \ifthenelse{\boolean{articletitles}}{{\it {Measurement of $b$ hadron
  production fractions in 7 TeV $pp$ collisions}},
  }{}\href{http://dx.doi.org/10.1103/PhysRevD.85.032008}{Phys.\ Rev.\  {\bf
  D85} (2012) 032008}, \href{http://arxiv.org/abs/1111.2357}{{\tt
  arXiv:1111.2357}}\relax
\mciteBstWouldAddEndPuncttrue
\mciteSetBstMidEndSepPunct{\mcitedefaultmidpunct}
{\mcitedefaultendpunct}{\mcitedefaultseppunct}\relax
\EndOfBibitem
\bibitem{LHCb-PAPER-2012-037}
LHCb collaboration, R.~Aaij {\em et~al.},
  \ifthenelse{\boolean{articletitles}}{{\it {Measurement of the fragmentation
  fraction ratio $f_s/f_d$ and its dependence on $B$ meson kinematics}},
  }{}\href{http://dx.doi.org/10.1007/JHEP04(2013)001}{JHEP {\bf 04} (2013)
  001}, \href{http://arxiv.org/abs/1301.5286}{{\tt arXiv:1301.5286}}\relax
\mciteBstWouldAddEndPuncttrue
\mciteSetBstMidEndSepPunct{\mcitedefaultmidpunct}
{\mcitedefaultendpunct}{\mcitedefaultseppunct}\relax
\EndOfBibitem
\bibitem{LHCb-CONF-2013-011}
{LHCb collaboration}, \ifthenelse{\boolean{articletitles}}{{\it {Updated
  average $f_{s}/f_{d}$ $b$-hadron production fraction ratio for $7 \tev$ $pp$
  collisions}}, }{}
  \href{http://cdsweb.cern.ch/search?p={LHCb-CONF-2013-011}&f=reportnumber&action_search=Search&c=LHCb+Reports&c=LHCb+Conference+Proceedings&c=LHCb+Conference+Contributions&c=LHCb+Notes&c=LHCb+Theses&c=LHCb+Papers}
  {{LHCb-CONF-2013-011}}\relax
\mciteBstWouldAddEndPuncttrue
\mciteSetBstMidEndSepPunct{\mcitedefaultmidpunct}
{\mcitedefaultendpunct}{\mcitedefaultseppunct}\relax
\EndOfBibitem
\bibitem{LHCb-PAPER-2011-035}
LHCb collaboration, R.~Aaij {\em et~al.},
  \ifthenelse{\boolean{articletitles}}{{\it {Measurement of $b$-hadron
  masses}}, }{}\href{http://dx.doi.org/10.1016/j.physletb.2012.01.058}{Phys.\
  Lett.\  {\bf B708} (2012) 241}, \href{http://arxiv.org/abs/1112.4896}{{\tt
  arXiv:1112.4896}}\relax
\mciteBstWouldAddEndPuncttrue
\mciteSetBstMidEndSepPunct{\mcitedefaultmidpunct}
{\mcitedefaultendpunct}{\mcitedefaultseppunct}\relax
\EndOfBibitem
\bibitem{CDFMasses}
CDF collaboration, \ifthenelse{\boolean{articletitles}}{{\it {Measurement of
  $b$ hadron masses in exclusive $J/\psi$ decays with the CDF detector}},
  }{}\href{http://dx.doi.org/10.1103/PhysRevLett.96.202001}{Phys.\ Rev.\ Lett.\
   {\bf 96} (2006) 202001}, \href{http://arxiv.org/abs/hep-ex/0508022}{{\tt
  arXiv:hep-ex/0508022}}\relax
\mciteBstWouldAddEndPuncttrue
\mciteSetBstMidEndSepPunct{\mcitedefaultmidpunct}
{\mcitedefaultendpunct}{\mcitedefaultseppunct}\relax
\EndOfBibitem
\bibitem{Hiller:2013cza}
G.~Hiller and R.~Zwicky, \ifthenelse{\boolean{articletitles}}{{\it
  {(A)symmetries of weak decays at and near the kinematic endpoint}},
  }{}\href{http://dx.doi.org/10.1007/JHEP03(2014)042}{JHEP {\bf 03} (2014)
  042}, \href{http://arxiv.org/abs/1312.1923}{{\tt arXiv:1312.1923}}\relax
\mciteBstWouldAddEndPuncttrue
\mciteSetBstMidEndSepPunct{\mcitedefaultmidpunct}
{\mcitedefaultendpunct}{\mcitedefaultseppunct}\relax
\EndOfBibitem
\end{mcitethebibliography}
\ifx\mcitethebibliography\mciteundefinedmacro
\PackageError{LHCb.bst}{mciteplus.sty has not been loaded}
{This bibstyle requires the use of the mciteplus package.}\fi
\providecommand{\href}[2]{#2}

\newpage                                                                                                            
\centerline{\large\bf LHCb collaboration}
\begin{flushleft}
\small
R.~Aaij$^{41}$, 
B.~Adeva$^{37}$, 
M.~Adinolfi$^{46}$, 
A.~Affolder$^{52}$, 
Z.~Ajaltouni$^{5}$, 
S.~Akar$^{6}$, 
J.~Albrecht$^{9}$, 
F.~Alessio$^{38}$, 
M.~Alexander$^{51}$, 
S.~Ali$^{41}$
% ...

{\footnotesize \it
% ...
$ ^{5}$Clermont Universit\'{e}, Universit\'{e} Blaise Pascal, CNRS/IN2P3, LPC, Clermont-Ferrand, France\\
$ ^{6}$CPPM, Aix-Marseille Universit\'{e}, CNRS/IN2P3, Marseille, France\\
% ...
$ ^{9}$Fakult\"{a}t Physik, Technische Universit\"{a}t Dortmund, Dortmund, Germany\\
% ...
$ ^{37}$Universidad de Santiago de Compostela, Santiago de Compostela, Spain\\
$ ^{38}$European Organization for Nuclear Research (CERN), Geneva, Switzerland\\
% ...
$ ^{41}$Nikhef National Institute for Subatomic Physics, Amsterdam, The Netherlands\\
% ...
$ ^{46}$H.H. Wills Physics Laboratory, University of Bristol, Bristol, United Kingdom\\
% ...
$ ^{51}$School of Physics and Astronomy, University of Glasgow, Glasgow, United Kingdom\\
$ ^{52}$Oliver Lodge Laboratory, University of Liverpool, Liverpool, United Kingdom\\
% ...
}
\end{flushleft}

\end{document}